\documentclass[preprint,12pt]{elsarticle}

\usepackage{amssymb}
\usepackage{amsmath}
\usepackage[english]{babel}

\usepackage{tabularx}
\usepackage{textgreek}

\usepackage[usenames]{color}

\usepackage{verbatim}

\journal{Chaos, Solitons and Fractals}

\date{November 10, 2026}

\begin{document}

\begin{frontmatter}
%Асимметрия порогов и пространственная самоорганизация активности в двумерной модели нейроглиоваскулярной единицы 
% \title{
% Cell-type specific dynamics and activity escape from depolarization block in a two-dimensional neurogliovascular model}

%\title{The race to avoid depolarization block by pyramidal neurons in a spatial neurogliovascular unit model}

\title{Self-organized fractional silencing of pyramidal neurons under global depolarization in a spatial neuron-glial microcircuit model}

\tnotetext[0]{This manuscript is a preprint and has not been peer-reviewed. It is currently under consideration for publication in \textit{Chaos, Solitons \& Fractals}.}

\author[1]{Stanislav V. Lagosha} %% Author name
\affiliation[1]{organization={Department of Biophysics, Biological Faculty, Lomonosov Moscow State University},%Department and Organization
            addressline={Leninskie Gory, 1/24}, 
            city={Moscow},
            postcode={119234}, 
            %state={},
            country={Russia}}
\author[2]{Andrey Yu. Verisokin\corref{cor1}} %% Author name
\ead{ffalconn@mail.ru }

\author[2]{Darya V. Verveyko} %% Author name

\affiliation[2]{organization={Department of Theoretical Physics, Kursk State University},%Department and Organization
            addressline={Radishcheva st., 33}, 
            city={Kursk},
            postcode={305000}, 
            %state={},
            country={Russia}}

\author[1]{Alyona V. Berezhnaya} %% Author name

\author[1]{Nadezda A. Brazhe} %% Author name

\author[1]{Alexey R. Brazhe\corref{cor1}} %% Author name
\ead{brazhe@biophys.msu.ru}
% \affiliation[3]{organization={Department of Molecular Neurobiology, Institute of Bioorganic Chemistry RAS},%Department and Organization
%             addressline={Miklukho-Maklaya st., 16/10}, 
%             city={Moscow},
%             postcode={117997}, 
%             %state={},
%             country={Russia}}
\cortext[cor1]{Corresponding authors}

%Unlike interneurons, which exit DB only passively after ionic homeostasis is restored, pyramidal neurons can actively recover spiking under sustained depolarization. 
%The exit from the depolarized state is driven by the slow inactivation of voltage-gated calcium currents, which reduces the steady-state membrane depolarization and reactivates fast sodium channels. 

%% Abstract
\begin{abstract}
Pathological transitions in nervous tissue, including epileptiform discharges and cortical spreading depolarization, depend on the interplay between cellular and interstitial mechanisms. We develop a biophysically detailed two-dimensional model of the neurogliovascular unit (NGVU) that integrates neuronal populations in the Hodgkin-Huxley formalism, astrocyte-regulated extracellular potassium dynamics, extracellular neurotransmitter spillover, and noradrenergic neuromodulation. Numerical simulations show that inhibitory interneurons fall into depolarization block (DB) at significantly lower extracellular potassium concentrations and stimulation intensities compared to pyramidal neurons. Early loss of inhibition shifts the local excitation-inhibition balance and triggers network hyperexcitability. 

In the spatially extended network, exceeding a critical stimulation threshold induces a transition from heterogeneous but generally spatially uniform firing to the fragmentation of the network into stable active microclusters with self-sustained high-frequency firing. Pyramidal neurons with lower excitatory synaptic input receive less recurrent excitation, which protects them from entering DB. By continuously releasing potassium ions and glutamate into the extracellular space, pyramidal neurons can cause strong depolarization in their more densely connected neighbors for an extended period of time and thus push them into a state of deep DB.

Simulations of localized astrocytic dysfunction predict a spatial limit for potassium clearance defects. Small damaged cores are fully compensated by diffusion and uptake by surrounding healthy astrocyte units. But if the size of the damage is large enough, the pathological activity-dependent buildup of potassium and firing positive feedback loop disrupts homeostasis mechanisms in the surrounding units and promotes the formation of persistent depolarized centers with self-sustained, spatially segregated pathological activity.

\end{abstract}
% -- AB: 13:55 09.11 абстракт посмотрел
% а Стас уже поменял...


%%Research highlights -- ОК (AB)
\begin{highlights}

\item Interneurons enter depolarization block at lower current and potassium thresholds than pyramidal cells.

\item Early inhibitory shutdown shifts the E/I balance, triggering runaway excitation.   

\item Strong depolarizing input leads to self-organized segregation of rapidly firing and silent pyramidal cells instead of full depolarization block.

\item Network topology and potassium clearance govern spatial pattern formation.

\item Exceeding the spatial scale of metabolic defects breaks tissue homeostasis.

\end{highlights}

%% Keywords

\end{frontmatter}

%\linenumbers


\section{Introduction}
\label{sec1}

%О чем мы?
%Роль гомеостаза в NGVU --- Нормальное функционирование мозга требует управления взаимосвязей соседних нейронов. Концепция NGVU объединяет динамику нейронов, метаболическую поддержку со стороны астроцитов и кровоток в единую функциональную систему.

The concept of the neurogliovascular unit (NGVU) in theoretical neurobiology reflects the shift from the analysis of isolated excitable elements to more holistic modeling of the neural tissue as a coupled, multi-component system~\cite{verkhratsky2018physiology, baldwin2024astrocyte}. Experimental evidence suggests that stable activity of local neuronal ensembles is impossible without continuous metabolic support from astrocytes and adequate blood perfusion provided by the vascular network.
The NGVU can be therefore considered a minimal functional unit of nervous tissue, which coordinates electrophysiological dynamics with local metabolic demand and supply~\cite{zhu2022neurovascular, verisokin2023computational}.
Physiological stability of neural dynamics requires maintenance of ionic composition of the interstitial space within relatively narrow bounds. On the other hand, accumulation of potassium ($\text{K}^+$) driven by synaptic activity acts as a modulator of neuronal excitability. Altered extracellular $\text{K}^+$ clearance shifts the resting membrane potential to depolarization and can trigger pathological activity~\cite{somjen1979extracellular}. The primary role in keeping $\text{K}^+$ balance and clearing neurotransmitters is played by astrocyte network, where individual cells are coupled via gap junctions into a distributed system of spatial buffering. 
%syncytium is our model assumption, maybe we should omit it from introduction astrocytic syncytium->astrocyte network
Thus, astrocytes modulate baseline excitation-inhibition balance and maintain both ionic homeostasis and metabolic stability under varying levels of neuronal activity~\cite{verkhratsky2018physiology}. Astrocytes exert control over interstitial composition through potassium siphoning by a spectrum of potassium channels, glutamate clearance by EAAT2 transporters~\cite{verkhratsky2018physiology}, GABA clearance or release via GAT transporters~\cite{wu2007nonvesicular, boddum2016astrocytic} and other mechanisms.

Specific mechanisms of spontaneous spatiotemporal pattern formation during pathological network activity remain a subject of active theoretical research, combining detailed computational modeling~\cite{ursino2006travelling, destexhe2009self} with more abstract mathematical frameworks~\cite{kramer2005pathological}. Spatial pattern formation has also been studied in networks with non-standard neuronal coupling schemes~\cite{fateev2024chimera}. 
One of the open questions is how differences in activation thresholds across neuronal populations govern transitions into collective depolarization block (DB) under sustained depolarization or elevated extracellular $\text{K}^+$ conditions. 
Understanding biophysical mechanisms of loss and recovery of normal activity in distributed NGVU networks~\cite{verisokin2025computational} is essential for unraveling the mechanisms of neuropathologies such as epilepsy and cortical spreading depolarization (CSD)~\cite{wei2014unification, ullah2015role}.

%Стоит отметить, что механизмы спонтанного формирования пространственно-временных структур при патологическом функционировании сети до сих пор остаются предметом активных теоретических исследований с использованием как моделирования ~\cite{ursino2006travelling, destexhe2009self}, так и более абстрактных математических построений ~\cite{kramer2005pathological}. Без внимания также не остаётся и исследование формирования пространственных патернов в сетях с нестандартным формализмом связей между нейронами ~\cite{fateev2024chimera}. Особый интерес представляет вопрос о том, как различие порогов активации у нейронов различных популяций определяет специфические переходы ткани в состояния коллективного деполяризационного блока. Понимание биофизической природы восстановления активности  в распределённых сетях NGVU~\cite{verisokin2025computational} имеет фундаментальное значение для описания этиологии нейропатологий, таких как эпилепсия и распространяющаяся кортикальная деполяризация~\cite{wei2014unification, ullah2015role}.

%Зачем мы?
%Нарушение ионного баланса (прежде всего концентрации внеклеточного калия) является основным фактором перехода ткани в патологические режимы, такие как эпилептическая активность или распространяющаяся депрессия.

Even minor deviations in extracellular $\text{K}^+$ concentration from its resting level can markedly alter the activity of neuronal populations and local neurotransmitter levels~\cite{dietz2023local}. Extracellular $\text{K}^+$ accumulation above threshold values (around 10--12~mM) is considered a trigger for a range of pathological cortical states~\cite{somjen1979extracellular,sawant2025pathophysiological}. Specifically, such shifts drive focal hyperexcitability underlying epileptiform discharges and, in extreme cases, initiate CSD, which is characterized by a profound suppression of electrical activity in the tissue~\cite{knowlton2022modeling}.

%Фундаментальное значение ионного баланса интерстициальной среды подтверждается тем фактом, что даже незначительные отклонения концентрации внеклеточного калия от уровня состояния покоя способны радикально изменять активность нейронных популяций, а также локальный уровень медиаторов ~\cite{dietz2023local}. Считается, что накопление калия выше пороговых значений (порядка 10–12 мМ) является пусковым механизмом для целого спектра патологических состояний коры головного мозга~\cite{somjen1979extracellular,sawant2025pathophysiological}. В частности, подобные сдвиги сопряжены с развитием фокальной гипервозбудимости, лежащей в основе эпилептиформных разрядов, а в критических случаях — с инициацией CSD, характеризующейся глубоким угнетением электрической активности ткани ~\cite{knowlton2022modeling}.

The transition to pathological dynamics is driven by a positive feedback loop between extracellular $\text{K}^+$ concentration and $\text{K}^+$ efflux, mediated by the functional state of the neurons. Rising extracellular $\text{K}^+$ causes sustained membrane depolarization, which initially lowers the action potential (AP) threshold and shifts the excitation-inhibition balance toward excitation, further destabilizing the network. However, at a certain level of depolarization, the inactivation of fast sodium channels becomes persistent, driving the neuron into a state of DB. Computational models of epilepsy show that the balance between synaptic excitation and the ionic dynamics of the microenvironment determines whether the network transitions into rhythmic activity or enters a phase of stationary electrical silence~\cite{wei2014unification,cressman2009influence,ullah2009influence}.

%Механизм патологизации динамики обусловлен положительной обратной связью между калием во внешней среде и его выбросом через функциональное состояние нейронов. Рост калия вызывает устойчивую деполяризацию мембран, что на начальных этапах приводит к снижению порога генерации потенциалов действия и смещению баланса торможения-возбуждения в сторону возбуждения, что в дальнейшем дестабилизирует сеть. Однако при достижении определенного уровня деполяризации инактивация быстрых натриевых каналов становится длительной, переводя нейрон в состояние деполяризационного блока (ДБ). В рамках вычислительных моделей эпилепсии показано, что именно баланс между синаптическим возбуждением и ионной динамикой микроокружения определяет, перейдёт ли сеть в режим ритмической активности или в фазу стационарного электрического молчания~\cite{abubakar2026beyond}.

The threshold nature of these transitions is of particular interest. Impaired astrocytic $\text{K}^+$ clearance caused by transporter dysfunction or extracellular space shrinkage (e.g., during cytotoxic edema), limits the compensatory capacity of the NGVU to dissipate excess ions, as observed in various central nervous system disorders~\cite{verkhratsky2018physiology,ullah2015role}. Under these conditions, local perturbations in the $\text{K}^+$ profile can expand spatially, forming persistent patterns of depolarized zones. Although the underlying ionic mechanisms are well documented~\cite{cressman2009influence, somjen2001ions}, the dynamical properties of the transition boundary into DB and the conditions for subsequent system relaxation in spatially extended networks remain poorly understood~\cite{ullah2009influence, dahlem2013transient, baspinar2025gabaergic}. Identifying the conditions under which nervous tissue can spontaneously restore homeostasis following DB episodes is necessary for understanding the progression of focal brain pathologies, such as ischemic penumbra and epileptic foci.

%Особый интерес представляет пороговый характер данных переходов. Нарушение клиренса калия астроцитами вследствие патологии транспортеров или уменьшения объема внеклеточного пространства (например, при отеке ткани) ограничивает компенсаторные возможности NGVU по диссипации избытка ионов, что обнаруживается при различных заболеваниях ЦНС ~\cite{zhu2022neurovascular}. В этих условиях локальные возмущения калиевого профиля могут масштабироваться, формируя устойчивые пространственные паттерны деполяризованных зон. Несмотря на широкое освещение ионных механизмов в литературе, динамические свойства границы перехода в режим блока и условия последующей релаксации системы в пространственно-протяженных сетях изучены фрагментарно [ссылки]. Определение условий, при которых ткань способна к самопроизвольному восстановлению гомеостаза после эпизодов ДБ, необходимо для глубокого понимания патогенеза таких структурных повреждений мозга, как ишемические поражения и фокальные эпилептические очаги.

%%%%%%%%%%%%%%%%%%%%%%%%%%%%%

% краткая выжимка содержания для конца Введения

At the single-cell level, excessive excitation terminates spiking, driving the neuron into a quasi-stationary depolarized state when it loses the ability to generate APs.
%+ AB: Мне вот это предложение не понятно. Pathological dynamics -- это, все-таки, когда нейрон генерирует спайки, а если он уходит в ДБ, то уже не генерирует. Поэтому сказать, что на уровне одиночной клетки переход к патологической динамике осуществляется через деполяризационный блок -- неверно.
% -- Если в абзаце только 1 предложение, то, как правило, разбиение на отдельные абзацы и не нужно
The primary mechanism underlying DB is the voltage-dependent inactivation of fast $\text{Na}^+$ channels that develops with sufficient membrane depolarization, which prevents channel recovery and renders the cell refractory to synaptic input. 
Under pathological conditions, the emergence of DB is triggered by critical extracellular $K^{+}$ accumulation. 
Inhibitory interneurons demonstrate higher sensitivity to depolarization block than excitatory pyramidal neurons, which leads to early termination of inhibition in the network and runaway excitation of pyramidal cells~\cite{ziburkus2006interneuron, karlocai2014physiological, baspinar2025gabaergic, aussel2022cell}.

Traditionally, the state of DB is interpreted as a stable, steady-state regime in the form of a passive pause in network activity without any intrinsic dynamics. The collective cessation of AP generation during critical potassium buildup represents the inactivation of neuronal electrogenic properties until interstitial ionic homeostasis is restored. %Consequently, DB was long regarded exclusively as a passive pause in network activity devoid of complex intrinsic dynamics.
However, recent biophysical findings and modeling demonstrate a fundamental possibility of spike activity recovery directly under sustained depolarizing pressure, driven by continuous ion concentration dynamics and membrane potential shifts~\cite{wei2014unification, ullah2015role}.
%Consequently, DB was long regarded exclusively as a passive pause in network activity devoid of complex intrinsic dynamics. However, recent biophysical findings indicate the fundamental possibility of spike activity recovery directly under sustained depolarizing pressure. Single-cell analyses of current–voltage characteristics highlight the role of slow conductances — specifically, voltage-gated calcium currents and the modulation of NMDA receptors — in initiating ``escape'' mechanisms from the block~\cite{knowlton2022modeling}. 
%By reducing steady-state depolarization via slow voltage-gated calcium currents, these mechanisms allow sodium channels to partially recover from inactivation.
%By reducing steady-state membrane depolarization and supporting ionic regulation via glial transport mechanisms, these mechanisms enable the partial reactivation of sodium channels.
%+ AB: конкретно у нас механизм другой же получился
Elucidating the conditions under which the tissue manages to avoid the terminal phase of deep DB is essential for understanding brain resilience under metabolic and electrical overload~\cite{knowlton2022modeling,abubakar2026beyond}. Nevertheless, the vast majority of existing theoretical studies remain limited to zero-dimensional (0D) point models, which can not account for the spatial component of the effects in neural tissue.
%precluding the study of systemic effects in real tissue.
In spatially extended systems, activity dropout and recovery are driven by collective network dynamics as well as diffusive interactions in the interstitial medium, while astrocytic clearance, and local synaptic connectivity distributions jointly shape propagation patterns and determine the spatial boundaries of pathological regimes~\cite{somjen1979extracellular,ullah2009influence}.

Nevertheless, how local competition of individual neurons for inactivation avoidance at microscopic level translates into macroscopic patterns of pathological self-organization remains an open question. 
In particular, it is unclear how local homeostatic breakdowns, such as impaired potassium clearance by astrocytes, or focal pathological depolarization determine whether a region recovers or gets locked in a sustained depolarized state.

In this paper, we address these gaps by developing a biophysically detailed spatial 2D model of NGVUs that integrates interacting neuronal populations, astrocytic homeostasis, and interstitial ion diffusion. Our work aims to elucidate the conditions governing neuronal recovery from DB, evaluate the impact of synaptic topology and $E/I$ heterogeneity on network fragmentation, and determine how local metabolic disruptions scale up to trigger large-scale pathological patterns akin to cortical spreading depolarization and focal epilepsy.

\section{The model}
%Stas - is the take about key components still true cause we still get N and G. maybe some of U cuase of extracellular medium.

\begin{figure}%[htbp]
  \begin{center}
\resizebox{1\columnwidth}{!}{
\includegraphics{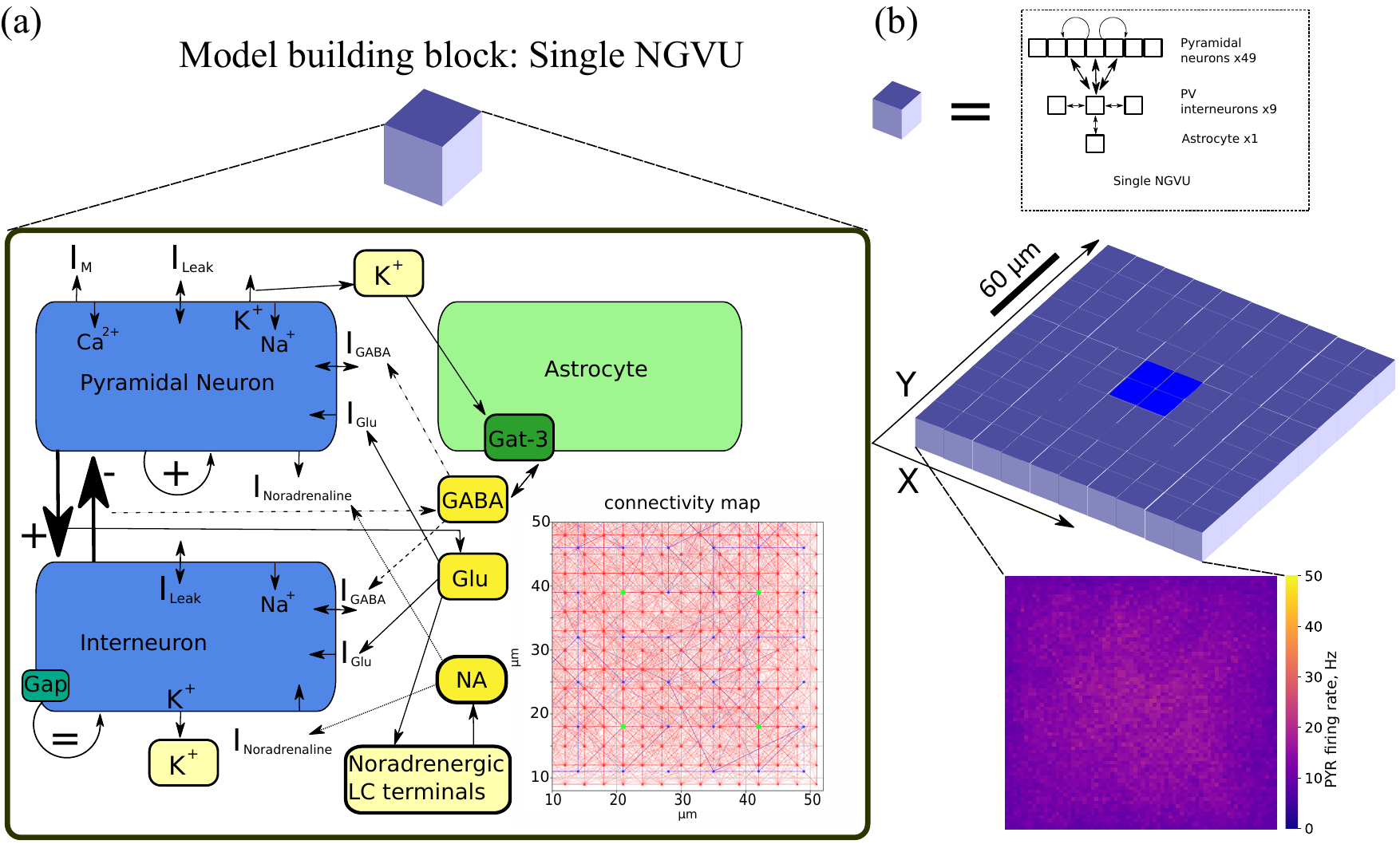}}
 \end{center}
\caption{Functional and topological schematic of the model. 
(a) Large rectangles: cell types. Yellow boxes: extracellular substances. Dark green boxes: functional proteins ('Gap' for potential-equalizing gap junctions; 'Gat-3' for the GABA transporter mediating thermodynamic equilibrium). Inter-compartment arrows indicate substance transport. Large arrows with ($+$) and ($-$) signs denote excitatory and inhibitory synapses, respectively. Bottom-right inset: connectivity map for the components of the simulated domain, where lines indicate the direction of connections between network nodes: astrocytes (green dots), pyramidal neurons (red dots), and PV interneurons (blue dots).
(b) Two-dimensional structure of the solution domain, where individual cells represent single NGVUs comprising interconnected groups of an astrocyte, and sets of pyramidal neurons and PV interneurons; the bottom image displays an example of the firing rate distribution within the simulated domain.}
\label{Model_Scheme}
\end{figure}

%Based on the available experimental data, we developed a functional model of the NGVUs (see Fig.~\ref{Model_Scheme}(a)). \textcolor{yellow}{The model includes key NGVU components} and reproduces key mechanisms of their interactions and physiological responses to stimulation. Below, we describe the main mathematical components of the model. Baseline parameters and descriptions of all variables are provided in the Supplementary Material.
Based on available experimental data, we developed a functional model of the neurogliovascular unit (see Fig.~\ref{Model_Scheme}a).
Although the neurogliovascular framework generally encompasses hemodynamics, for the purposes of this study, vascular tone is treated as a constant homeostatic sink for ion and neurotransmitter clearance. Such a configuration allows for a focused analysis of the glio-interstitial and neuronal dynamics, which we identify as the primary drivers of the spatiotemporal patterns and homeostatic collapse analyzed here. The model reproduces the key mechanisms of interaction within this ensemble and its physiological responses to external stimulation. 
Below, we describe the main mathematical components of the model. Baseline parameters and descriptions of all variables are provided in the Supplementary Material.

%Stas regulating-> mediating

\subsection{Neurons}

The network consists of excitatory pyramidal neurons and inhibitory interneurons modeled using Hodgkin–Huxley formalism based on a detailed modeling study of human hippocampus~\cite{aussel2022cell}. % AB: Нам нужно сослаться, откуда берется такой набор токов
Pyramidal neurons include fast voltage-gated $\mathrm{Na}^+$ and $\mathrm{K}^+$ currents, slow $\mathrm{K}^+$ and $\mathrm{Ca}^{2+}$ currents, and a leak current. Interneurons include only the fast $\mathrm{Na}^+$, fast $\mathrm{K}^+$, and leak currents. We also include receptor-mediated currents: a $\mathrm{GABA}$-dependent $\mathrm{Cl}^-$ current and an NMDA-dependent cation % AB: checkme!
current activated by extracellular glutamate. These currents couple the neuronal models to the astrocyte subsystem.

The dynamics of the membrane potential of pyramidal neurons ($V$) is described by the following equation:
\begin{equation}
\label{eq:dV1}
\begin{aligned}
{C_{\mathrm{mem}}}{S_{\mathrm{pyr}}} \frac{dV}{dt}
={}& I_{\mathrm{stim}} - \Bigl( 
I_{\mathrm{Na}} + I_{\mathrm{K}} + I_{\mathrm{Ca}} + I_M \\ &\quad + I_{\mathrm{leak}} + I_{\mathrm{tonic}}
+  I_{\mathrm{Glu}} +I_{NA} + \sum I_{\mathrm{syn}} \Bigr),
\end{aligned}
\end{equation}
% AB: исправил знак у Istim -- классически он положительный
where $C_{\mathrm{mem}}$ is specific membrane capacitance, $S_{\mathrm{pyr}}$ is surface area, and $I_{\mathrm{stim}}$ is an applied bias current. The intrinsic currents include sodium ($I_{\mathrm{Na}}$), potassium ($I_{\mathrm{K}}$), calcium ($I_{\mathrm{Ca}}$), slow potassium ($I_M$), extracellular noradrenaline based current ($I_{NA}$) and leak ($I_{\mathrm{leak}}$) currents. 
% Повтор параграфа выше, но не страшно
The extrasynaptic and modulatory currents comprise $I_{\mathrm{tonic}}$ (GABA-mediated) and $I_{\mathrm{Glu}}$ (glutamate-mediated). Synaptic input consists of recurrent network connections ($\sum I_{\mathrm{syn}}$).% and stochastic excitatory drive from the entorhinal cortex ($I_{\mathrm{EC}}$), modeled as Poisson spike trains.

The ionic and receptor-mediated currents are defined as:
%Стас - убрал всё что связано с площадью из токов. оставил её только в конечном уравнении чтоб мы её хоть раз упомянули. все проводимости в ионных токах считаем что они уже умножены на неё.%
\begin{equation}
\label{eq:Flows1}
\begin{aligned}
I_{\mathrm{Na}}   &= \bar{g}_{\mathrm{Na}}  m^3 h (V - V_{\mathrm{revNa}}), \\
I_{\mathrm{K}}    &= \bar{g}_{\mathrm{K}}  n^4 (V - V_{\mathrm{revK}}), \\
I_{\mathrm{Ca}}   &= \bar{g}_{\mathrm{Ca}}  m_{\mathrm{Ca}}^2 h_{\mathrm{Ca}} (V - V_{\mathrm{revCa}}), \\
I_M               &= \bar{g}_M  p (V - V_{\mathrm{revK}}), \\
I_{\mathrm{leak}} &= g_{\mathrm{leak}}  (V - V_{\mathrm{revleak}}), \\
I_{\mathrm{tonic}}&= g_{\mathrm{\text{GABA}}}  (V - V_{\mathrm{rev\text{GABA}}}), \\
I_{\mathrm{Glu}}  &= g_{\mathrm{Glu}}  (V - V_{\mathrm{revGlu}}),\\
I_{\mathrm{NA}}  &= g_{\mathrm{NA}} (V - V_{\mathrm{revK}}),
%{I_{NA}} &= {v_{NA}}\frac{{{{[NA]}}}}{{{{[NA]}} + k_{NA}}}.\\ -- AB: это не тот!!!
% AB: А почему I_glu есть только у pyr?
\end{aligned}
\end{equation}
% Стас мы же убрали площадь тогда это просто максимальные проводимости
where $\bar{g}$ and $g$ are maximal conductances, and $V_{\mathrm{rev}}$ terms are the corresponding reversal potentials. The gating variables $m$, $n$, $m_{\mathrm{Ca}}$ and $p$ describe activation, while $h$ and $h_{\mathrm{Ca}}$ describe inactivation.

Unlike the other reversal potentials, $V_{\mathrm{revK}}$ varies with extracellular and intracellular potassium:
\begin{equation}
V_{\mathrm{revK}} = \frac{RT}{F} \ln  \frac{[\mathrm{K}^+]_{\mathrm{ext}}}{[\mathrm{K}^+]_{i,\mathrm{pyr}}} .
\end{equation}
Here, $R$ is the universal gas constant, $T$ is absolute temperature, $F$ is the Faraday constant. Flow directions for all currents are summarized in Fig.~\ref{Model_Scheme}. 

The synaptic and neuromodulatory conductances associated with GABA, glutamate, and noradrenaline (NA) are modeled using Hill-type activation kinetics to account for receptor binding saturation. For the noradrenergic system, for example, the concentration-dependent conductance $g_{\text{NA}}$ is defined as:
\begin{equation}
\label{eq:gna}
g_{\text{NA}} = \frac{g_{\text{NA,max}}}{1 + \left( \frac{K_{50,\text{NA}}}{[\text{NA}]} \right)^n}
\end{equation}
where $g_{\text{NA,max}}$ is the maximal conductance, $K_{50,\text{NA}}$ represents the concentration for half-maximal activation, and $n$ is the Hill coefficient.

Interneuron membrane potential evolves according to:
\begin{equation}
\label{eq:dV_IN}
C_{\mathrm{mem}} S_{\mathrm{pv}} \frac{dV}{dt} = -\left( I_{\mathrm{Na}} + I_{\mathrm{K}} + I_{\mathrm{leak}} + I_{\mathrm{tonic}} + I_{\mathrm{gap}}+ \sum I_{\mathrm{syn}} \right),
\end{equation}
where $S_{\mathrm{pv}}$ is cell surface area, and $I_{\mathrm{gap}}$ accounts for electrical coupling between interneurons. Individual currents are defined similarly to Equation\,\eqref{eq:Flows1}.
%\begin{equation}
%Стас тоже убрал площадь%
%\label{eq:Flows_IN}
%\begin{aligned}
%I_{\mathrm{Na}}   &= \bar{g}_{\mathrm{Na}} } m^3 h (V - %V_{\mathrm{revNa}}), \\
%I_{\mathrm{K}}    &= \bar{g}_{\mathrm{K}} } n^4 (V - %V_{\mathrm{revK}}), \\
%I_{\mathrm{leak}} &= g_{\mathrm{leak}} } (V - %V_{\mathrm{revleak}}), \\
%I_{\mathrm{tonic}}&= g_{\mathrm{GABA}} } (V - %V_{\mathrm{revGABA}}).
%\end{aligned}
%\end{equation}
% Stas added text about pv specific parametrization
%Gating variables ($m, h, n$) and parameters match the pyramidal neuron definitions, with interneuron specific parametrization and that $V_{\mathrm{revK}}$ uses the interneuron intracellular concentration $[\mathrm{K}^+]_{i,\mathrm{pv}}$. Extracellular ion concentrations are taken from \cite{verkhratsky2018physiology}.
The gating variables ($m, h, n$) follow the same mathematical formalism as for pyramidal neurons but utilize interneuron-specific parameter sets. In particular, the potassium reversal potential $V_{\text{rev,K}}$ is calculated based on the intracellular concentration $[\text{K}^+]_{i,\text{pv}}$. Reference extracellular ion concentrations are obtained from \cite{verkhratsky2018physiology}.

The dynamics of all gating variables ($x \in \{m, h, n, p\}$) are governed by:
\begin{equation}
\label{eq:gating}
\frac{dx}{dt} = \alpha_x(V) (1 - x) - \beta_x(V) x,
\end{equation}
with voltage-dependent rate functions $\alpha_x(V)$ and $\beta_x(V)$.

Spikes are registered when $V$ crosses a threshold $V_{\mathrm{th}}$, which triggers synaptic events and neurotransmitter release. Total synaptic current is given by:
\begin{equation}
\label{eq:Isyn}
I_{\mathrm{syn}} = g_{\mathrm{syn}} (V - V_{\mathrm{syn}}) \sum_i s_i,
\end{equation}
where $g_{\mathrm{syn}}$ is maximal synaptic conductance, $V_{\mathrm{syn}}$ is the reversal potential, and $s_i$ is the gating variable of input $i$, modeled with bi-exponential kinetics (rise time $\tau_r = 1/\alpha$, decay time $\tau_d = 1/\beta$). Synaptic parameters depend on cell pairing and follow \cite{wheeler2015hippocampome}.

At each presynaptic spike arrival ($t_{\mathrm{spike}}$), the synaptic variables are modified as in \cite{roth2009modeling}:
\begin{equation}
\label{eq:s0s1}
\begin{aligned}
s_0(t_{\mathrm{spike}}^+) &\rightarrow s_1(t_{\mathrm{spike}}^-), \\
%s_1(t_{\mathrm{spike}}^+) &\rightarrow s_\infty \left(1 - e^{-1/\tau_s}\right) + %s_1(t_{\mathrm{spike}}^-) e^{-1/\tau_s + \beta},\\
s_1(t_{\text{spike}}^+) &\rightarrow \left( s_\infty \left(1 - e^{-\Delta t / \tau_s}\right) + s_1(t_{\text{spike}}^-) e^{-\Delta t / \tau_s} \right) e^{\beta \Delta t},
\end{aligned}
\end{equation}
where superscripts $-$ and $+$ denote values immediately before and after the spike, $\Delta t$ represents the time step or the numerical integration and $\beta$
denotes the synaptic decay rate constant.

Each spike also triggers a synapse-specific amount of neurotransmitter spillover into the extracellular space, activating extrasynaptic receptors \cite{scanziani2000gaba}. Chemical connections comprise 
%AMPA/NMDA % AB: мне казалось, что в текущей формулировке мы только AMPA моделируем?
AMPAR-mediated excitation (pyramidal-to-pyramidal and pyramidal-to-interneuron) and GABAergic inhibition (interneuron-to-pyramidal and interneuron-to-interneuron). Interneurons are additionally coupled by electrical gap junctions.

Gap junctions are included only between nearest-neighbor interneurons \cite{fukuda2000gap}:
\begin{equation}
\label{eq:Igap}
I_{\mathrm{gap},i} = g_{\mathrm{gap}} \sum_{j,i\neq j} (V_i - V_j),
\end{equation}
where $j$ runs over adjacent cells, and $g_{\mathrm{gap}}$ is the junction conductance.

Neuronal spiking activity drives potassium efflux into the extracellular space in proportion to repolarizing currents \cite{somjen1979extracellular}. The single-spike concentration increment was scaled down relative to reported in vivo values to avoid runaway depolarization in the network. 
% AB: ну мы потом и вызываем "runaway depolarization"
%Similarly, each presynaptic event releases a fraction of glutamate or GABA into the extracellular compartment to account for neurotransmitter spillover.

%Extracellular potassium couples the neuronal, astrocytic and diffusion compartments: shifts in $[\mathrm{K}^+]_{\mathrm{ext}}$ alter the potassium reversal potential, directly modulating cell excitability and astrocytic uptake.

\subsection{Astrocyte-controlled extracellular environment}

The extracellular microenvironment couples the cellular compartments via dynamic fluctuations in potassium [K$^+$]$_\text{e}$, glutamate $[\mathrm{Glu}]$ and GABA $[\mathrm{GABA}]$. 
Extracellular potassium sets the resting potentials of both neurons and astrocytes, with baseline clearance implicitly capturing $\mathrm{Na}^+/\mathrm{K}^+$-ATPase activity.
The equation for potassium has the following form:
\begin{equation}
    \frac{d[K^+]_e}{dt} = \frac{[K^+]_{bath} - [K^+]_e}{\tau_K}
    + \sum_{j_\text{PYR,PV}}\kappa_\text{K}\delta(t-t_{j})
    ,
\label{eq:potassium}
\end{equation}
where $\tau_K$ represents the characteristic clearance time constant, $j_{\text{PYR,PV}}$ spikes of any neuron located in the given NGVU with spike time $t_j$,  $\kappa_K$ is a fixed spike-related increase in [K$^+$]$_\text{e}$,  and $\delta$ is Dirac delta function. 
% The term $I_{pump}$ describes the active outward current of the sodium-potassium pump depending on the intracellular sodium and extracellular potassium concentrations, and the availability of intracellular metabolic resources. In the current model, the ATP concentration is assumed to be constant for simplicity. The pump dynamics are defined as follows:
% \begin{equation}
% \begin{aligned}
%     I_{pump} = \frac{2}{\left(1 + e^{-14 ([ATP]_{base} - 0.55)}\right) \left(1 + e^{3.5 - [K^+]_e}\right) \left(1 + e^{\frac{25 - [Na^+]_i}{3}}\right)}
% \end{aligned}
% \end{equation}

Glutamate and GABA govern synaptic and extrasynaptic conductances in neurons and activate glial transporters. Ambient glutamate activates high-affinity extrasynaptic NMDA receptors, contributing to tonic depolarization and modulating network excitability \cite{meur2007tonic}: 
\begin{equation}
    \frac{d[\text{Glu}]}{dt} = \frac{[\text{Glu}]_{amb} - [\text{Glu}]}{\tau_{\text{Glu}}}
    + \sum_{j_\text{PYR}}\kappa_\text{Glu}\delta(t-t_{j})
    . %+% k_{pot} \mathcal{F}([K^+]_e) +    S_{Glu}(t)
    \label{eq:glu}
\end{equation}
Here glutamate spillover is linked to synaptic activity of pyramidal neurons in a similar way to~\ref{eq:potassium}.

Extracellular [GABA] similarly drives a tonic, chloride-dependent conductance in both pyramidal neurons and interneurons. [GABA] concentrations are determined by the interplay between astrocytic GAT3-mediated uptake ($J_{in}$) and release ($J_{out}$) fluxes and spillover driven by PV interneuron synaptic activity:
\begin{equation}
    \frac{d[\text{GABA}]}{dt} = \sigma \left( J_{out} - J_{in} \right)
    + \sum_{j_\text{PV}}\kappa_\text{GABA}\delta(t-t_j)
    .
    \label{eq:gaba} 
    % + S_{GABA}(t)
\end{equation}
Driven by local electrochemical gradients, this dynamic flux balance establishes a mechanistic
link between neural network tonic inhibition, potassium homeostasis, and the glial membrane potential. Additional terms are omitted from the equations, as the clearance mechanism inherently account for the gradient dependence.

Also we account the currents induced by norepinephrine in neurons, mediated by the activation of the corresponding adrenergic receptors~\cite{pang1987differential}.
The model considers the effect of norepinephrine (NA) on both the calcium dynamics of the astrocyte and the functioning of the NGVU as a whole, through the effect of NA on the production of [IP$_3$], and its direct action on neuron through adrenoreceptors.

Equations governing norepinephrine dynamics, alongside current induced by [NA] concentration in neurons:
%Стас- член НА-диф по сути должен описывать диффузию норадреналина, но у нас есть она в явном виде, в итоге это просто член добавляющий некую постоянную прибавку к изменению норадреналина. Так как она на два порядка меньше {{[NA]}_{amb}} предлагаю его просто убрать из кода и из уравнений% убрал

\begin{equation}
\label{eq:NA}
\frac{{d\text{[NA]}}}{{dt}} =\frac{[\text{Glu}] - [\text{Glu}]_{\text{amb}}}{r_{\text{NG}}} + \frac{[\text{NA}]_{\text{amb}} - [\text{NA}]}{\tau_{\text{NA}}}%+ !!!N{A_{diff}}!!! .
\end{equation}

Spatial communication in the model is implemented via the lateral diffusion of $[K^+]_e$, $[\text{Glu}]$, $[\text{GABA}]$ and $[\text{NA}]$ (see Jupyter notebooks in the Supplementary Material). 
This diffusive transport couples the dynamics of adjacent NGVUs, allowing localized ionic and neurotransmitter fluctuations to propagate across the model space.
Note that this framework for neurotransmitter dynamics enables straightforward integration of additional neuromodulatory systems (e.g., serotonin or acetylcholine).

\section{Network topology and simulation details}

Simulations were performed on a two-dimensional grid of coupled NGVUs (Fig.\,\ref{Model_Scheme}b). Each unit comprises a single astrocyte together with local populations of pyramidal neurons and interneurons embedded in a shared extracellular space. Neurons are placed on regular 2D lattices with connectivity defined by inter-soma distances.

%Each NGVU occupies a $21 \times 21\,\mathrm{\mu m^2}$ square, matching typical astrocytic domain sizes \cite{baldwin2024astrocyte}. Within this area, pyramidal cells and interneurons are arranged on square lattices with grid spacings of $3\,\mathrm{\mu m}$ and $7\,\mathrm{\mu m}$, respectively. This yields 49 excitatory and 9 inhibitory neurons per domain (with interneurons comprising $\approx 15.5\%$ of the population), consistent with cortical excitation–inhibition proportions \cite{wilson1972excitatory, rotstein2020neurons} (Fig.\,\ref{Model_Scheme}b, upper panel).
% -- AB: вот здесь немного скользкий момент, модель гиппокампальная, а пропорции из коры?! Рецензенты могут потребовать прояснить этот вопрос.

Each NGVU occupies a $21 \times 21\,\mathrm{\mu m^2}$ square, matching typical astrocytic domain sizes. Within this area, pyramidal cells and interneurons are arranged on square lattices with grid spacings of $3\,\mathrm{\mu m}$ and $7\,\mathrm{\mu m}$, respectively (Fig.\,\ref{Model_Scheme}b, upper panel). This yields 49 excitatory and 9 inhibitory neurons per domain (with interneurons comprising $\approx 15.5\%$ of the population), consistent with anatomical and physiological proportions of local hippocampal and cortical networks~\cite{wheeler2015hippocampome, wilson1972excitatory, rotstein2020neurons}.

%
%This yields 49 excitatory and 9 inhibitory neurons per domain (pyramidal:interneuron ratio $\approx 5.4:1$, with interneurons comprising $\approx 15.5\%$ of the population), consistent with cortical excitation--inhibition proportions \cite{wilson1972excitatory, rotstein2020neurons} (Fig.\,\ref{Model_Scheme}b, upper panel).

Synaptic connections are assigned stochastically with connection probability decaying as a Gaussian function with Euclidean distance from the presynaptic neuron $p \sim e^{-\frac{d^2}{2\sigma^2}}$ (see right bottom inset in Fig.\,\ref{Model_Scheme}a, line thickness indicates connection probability). We used $\sigma=7$\,$\mu$m as baseline parameter and varied it from $0.2\times$ to $3\times$ of its original value in Fig.\,\ref{fig:coupled-connectivity}b--c.
%*** AB: нужно указать константу длины, с которой падает вероятность образовать связь, и желательно ее обосновать ссылкой на литературу! Я предполагаю, что мы именно ее меняли на рис. 4, поэтому нам также надо это указать здесь (диапазон, в которм мы ее могли менять. Также именно тут нужно указать, как делался full scaling. 
% 

This spatial layout couples local synaptic circuits, astrocytic transport and extracellular diffusion. Example firing rate dynamics of pyramidal neurons are shown in the bottom panel of Fig.\,\ref{Model_Scheme}b.

Simulations were implemented in Python using the Brian 2 simulator \cite{Stimberg:2019}. Full model equations, parameter sets, and run scripts are provided as Jupyter notebooks in the Supplementary Material and will be archived on Zenodo upon publication.
% -AB: выделение красным можно убрать   

% проверить цвета!

%red colour denotes the area of low-amplitude sub-threshold oscillations instead of action potentials (APs); purple indicates depolarization block; pink represents sustained firing of APs; orange indicates bursting activity; dark blue shows no reaction to stimulation; and dark purple represents a single AP at the onset of stimulation.

%%%%%%%%%%%%%%%%%%%%%%%%%%%%%%%%%%%%%%%%%%
\section{Simulation results}

\subsection{Electrophysiological characteristics and dynamics of the depolarization block}

% To elucidate biophysical mechanisms underlying the formation of pathological spatiotemporal patterns in neural tissue, we first investigated the dynamics of isolated pyramidal neurons (PYR) and inhibitory interneurons (PV) under varying stimulating current $I_{\mathrm{stim}}$ and extracellular potassium concentration $[K^+]_{e}$.
% -> rephrasing from QWEN:
To dissect the biophysical origins of pathological spatiotemporal patterns in neural tissue, we first characterized the dynamics of individual pyramidal neurons (PYR) and parvalbumin-positive (PV) interneurons in isolation, systematically varying the applied current $I_{\mathrm{stim}}$ and extracellular potassium concentration [K$^+$]$_\text{e}$.

%The two neuron types exhibit dissociate resistance to DB under increasing depolarizing current stimulation, as illustrated in Fig.\,\ref{fig2}a by examples of the membrane potential $V_m$ dynamics for a pyramidal neuron (red traces) and an interneuron (blue traces) in response to a step depolarizing stimulus switched on at $t = 0$\,s.

\begin{figure}%[htbp]
  \begin{center}
\resizebox{1\columnwidth}{!}{
\includegraphics{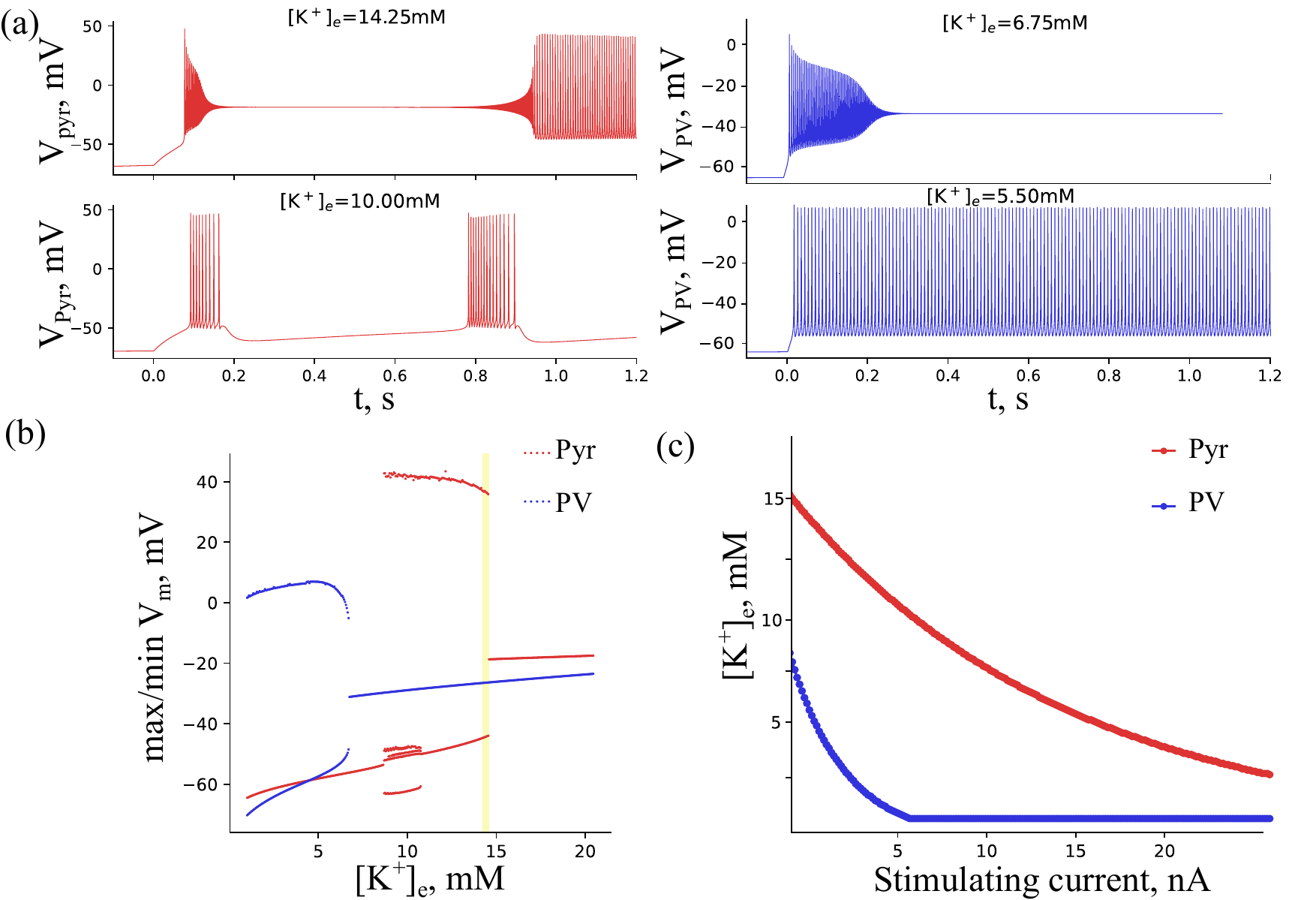}}
 \end{center}
\caption{Electrophysiological characteristics and activity regimes. (a) Examples of membrane potential $V_{m}$ dynamics for a pyramidal neuron (red) and an interneuron (blue) in response to depolarizing current steps with different amplitudes starting at $t=0$\,s. Note the bursting activity of pyramidal neurons and transition to depolarization block in the interneuron. 
(b) Bifurcation diagrams that local minima and maxima of potential values sampled from a 20-second simulation, plotted as a function of the extracellular potassium concentration ([K$^+$]$_\text{e}$); the vertical yellow line marks the narrow boundary of the post-DB recovery regime prior to terminal DB in the pyramidal neuron.
(c) Bifurcation parameter curves separating the firing and DB regimes in the $([\text{K}^+]_\text{e}, I_{\text{stim}})$ plane for both neuron types.
}
\label{fig2}
\end{figure}

Both neuron types display a range of activity regimes in response to long depolarizing current pulses and elevated [K$^+$]$_\text{e}$ as illustrated in Fig.\,\ref{fig2}a. Here, we exemplify the ability of PYR neurons to generate bursts of activity due to slow inactivation and deinactivation of calcium current and its ability to exit depolarization block (DB) that exists in a narrow parameter range. PV neurons, in contrast, show a typical fast-spiking response at low [K$^+$]$_\text{e}$ and stable exit to DB at higher [K$^+$]$_\text{e}$. 
%todo: добавить на рис. $[K^+]_{e}$, а то непонятно. 

Bifurcation analysis shows that PV neurons are more susceptible to depolarizing block (DB), entering this state at lower depolarizing currents and [K$^+$]$_\text{e}$ values than PYR neurons.
As seen from the bifurcation diagram with [K$^+$]$_\text{e}$ as the control parameter
%which shows the span of maximum and minimum membrane potential values ($\max V_m$, $\min V_m$) as a function of $[K^+]_{e}$ % -- это лучше в подрисуночную подпись (checkme: проверить, что есть)
(Fig.\,\ref{fig2}b), the transition from spiking to DB goes through a classical Andronov--Hopf bifurcation for both neuron types: there are well-defined threshold boundaries for the transition into DB. %and the recovery of AP generation. -- А где recovery?
The convergence point of the upper ($\max V_m$) and lower ($\min V_m$) potential branches into a single line corresponds to the critical potassium concentration at which the neuron loses excitability and enters DB. However, in pyramidal neurons, there is a narrow band before the terminal DB, where the neuron transiently enters DB, but the spiking activity can recover due to $I_{\text{Ca}^{2+}}$ inactivation as shown in (Fig.\,\ref{fig2}a). Interestingly, in interneurons, the transition to DB occurs at a dramatically lower [K$^+$]$_\text{e}$ concentration than in pyramidal neurons. Furthermore, as indicated by the DB bifurcation boundaries in Fig.\,\ref{fig2}c, this shift is also achieved at significantly lower stimulating current values.
The split of the minimum potential curve ($\min V_m$) of the pyramidal neuron into multiple levels at $[\text{K}^+]_\text{e} \approx 9\text{--}11\text{ mM}$ reflects the bursting activity regime, in which the local membrane potential minima within a burst and during inter-burst intervals differ in depth. Thus, the early functional shutdown of PV interneurons in comparison to PYR neurons
%combined with the capacity of pyramidal cells for recovery from DB,  %-- alas, this is not entirely true:)
creates a threshold disparity that forms the cellular basis of tissue spatial self-organization.

\subsection{Dissociation of activity in the potassium-uncoupled network}

\begin{figure}
    \centering
    \includegraphics[width=1\columnwidth]{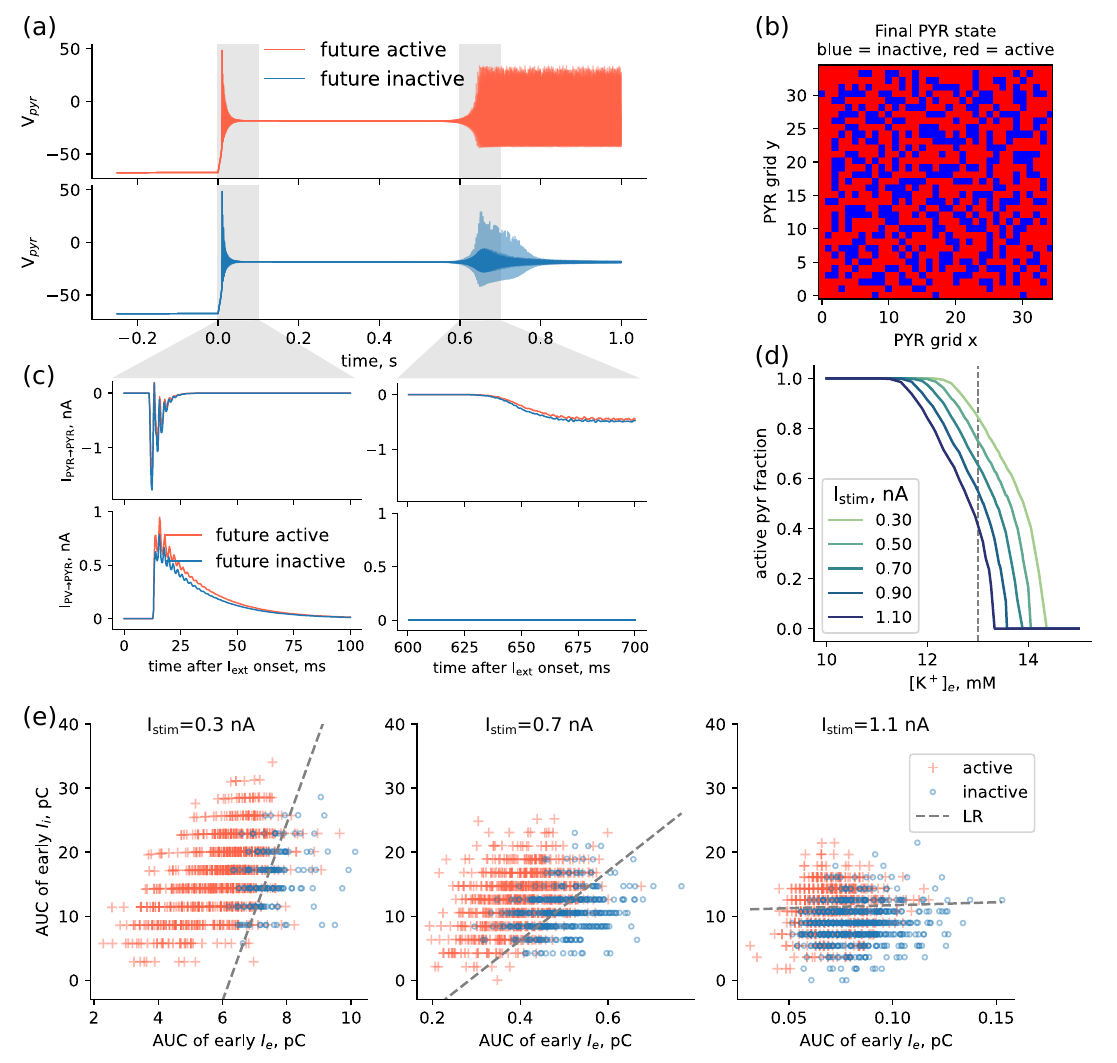}
    \caption{Segregation of active and inactive pyramidal neurons in the potassium-uncoupled model.
    (a) Example voltage traces for 10 randomly chosen active and 10 randomly chosen inactive neurons at [K$^+$]$_e$ = 13\,mM and $\text{I}_\text{stim}=0.7$\,nA. Transient silent period is due to slow inactivation of $\text{I}_{\text{Ca}}$. Shaded time windows are expanded in (c).
    (b) Final state of pyramidal neurons for the simulation shown in (a), active neurons shown in red, silent neurons are shown in blue.
    (c) Average excitatory (PYR$\rightarrow$PYR) and inhibitory (PV$\rightarrow$PYR) currents in the pyramidal neurons that will remain active (red) or become inactive (blue) for the time windows shaded in (a). Note that in the second time window all PV neurons are in the state of DB and the inhibitory current is zero.
    (d) Fraction of active pyramidal neurons as a function of  [K$^+$]$_e$ for different amplitudes of stimulation current $\text{I}_\text{stim}$. Vertical dashed line corresponds to [K$^+$]$_e$ = 13\,mM, the value used in (a--c) and (e). 
    (e) Separation of active and inactive neurons by the early integral inhibitory and excitatory currents, taken in the first time window shown in (c). Red markers --- active neurons, blue markers --- inactive, dashed line: separation by logistic regression. 
    %
    % AB:  future todo: надо было показать и активнсть тормозных нейронов, что они уходятв ДБ.
    %      или добавить общий растр как отдельный рисунок
    %
    }
    \label{fig:uncoupled}
\end{figure}

To understand how the excitability properties of isolated neurons define the network behavior under high depolarization drive and to delineate these effects from glia-based feedback loops, we first looked at activity of synaptically connected PYR and PV neurons distributed on a spatial grid with extracellular potassium concentration as a control parameter (Fig.~\ref{fig:uncoupled}). 
In the 10--15\,mM range of [K$^+$]$_\text{e}$, PV neurons easily switch to DB at minor excitation and do not provide any further inhibition in the network, allowing runaway excitation. 
Unexpectedly, within a range of [K$^+$]$_\text{e}$ and I$_\text{stim}$ values, not all PYR neurons undergo the DB bifurcation and thus some remain active, as shown in Fig.~\ref{fig:uncoupled}a--b for $[\text{K}^+]_\text{e}=13$\,mM and I$_\text{stim}=0.7$\,nA. 
Here, both subpopulations of neurons transiently enter DB, but then a fraction of neurons restore activity due to I$_\text{Ca}$ inactivation, while the other fraction remains in DB or recovers only transiently. 
We note, however, that without I$_\text{Ca}$ this activity dissociation would remain, just without the transient DB (data not shown). 
For a given $I_\text{stim}$ amplitude, the transition from fully active network to fully blocked occurs over a range of $>1$\,mM [K$^+$]$_\text{e}$ through the states with intermediate fractions of active neurons; increasing the depolarization current amplitude shifts this curve to the lower extracellular potassium concentrations~(Fig.\,\ref{fig:uncoupled}d).

Thus, within a range of specific I$_\text{stim}$ and [K$^+$]$_\text{e}$  combinations, some neurons go through the DB bifurcation and some do not. Because all PYR neurons in the potassium-uncoupled network experience the same extracellular potassium concentration and stimulation current, the only reason for the observed segregation must lie in their synaptic connectivity pattern and the synaptic currents they integrate. Indeed, the neurons that remain active tend to receive more inhibitory input during the transient activation of PV neurons in the first 100\,ms of stimulation as well as less excitatory input in the first  100\,ms and during the critical dissociation period~(Fig.\,\ref{fig:uncoupled}c). The relative importance of early excitatory and inhibitory currents changes with the amplitude of stimulation current: at low excitatory drive the separation is dominated by excitatory current while at high drive the separation depends on inhibitory current, as can be seen from the slopes of logistic regression lines~(Fig.\,\ref{fig:uncoupled}e).

%% ----------------------  Potassium- and NA-coupled model ------------------------------------
\subsection{Self-driven activity dissociation in the potassium-coupled network}

\begin{figure}
    \centering
    \includegraphics[width=1\columnwidth]{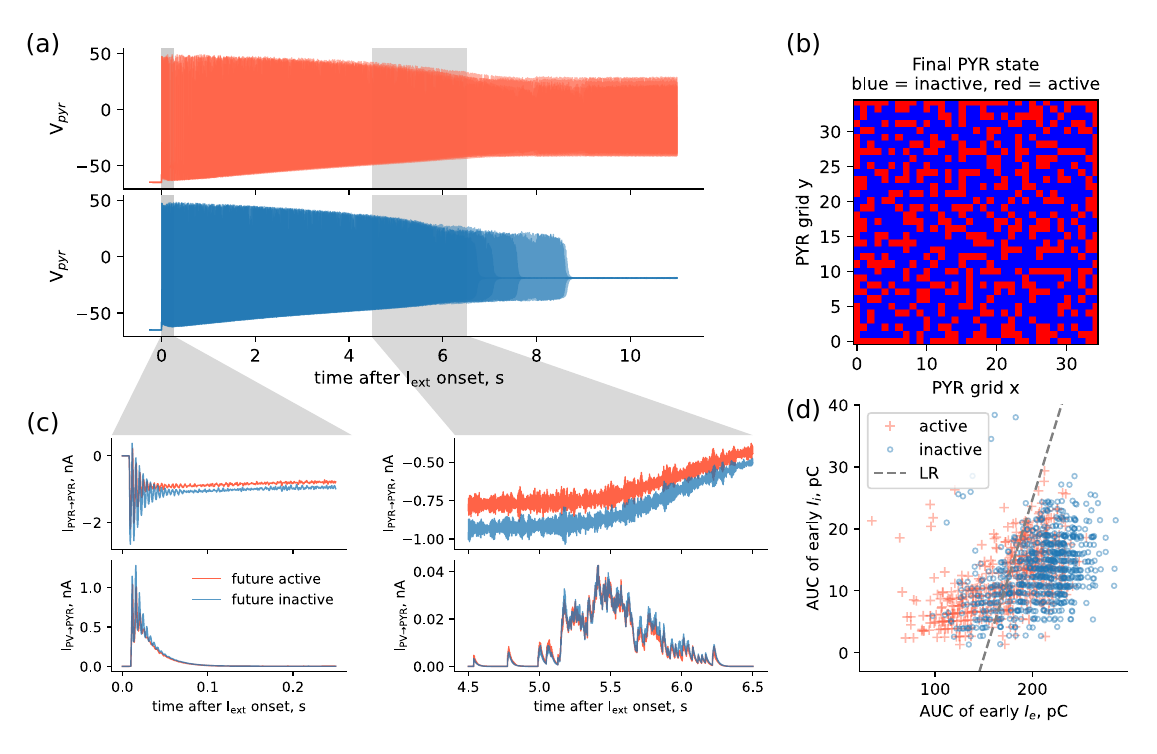}
    \caption{Segregation of active and inactive pyramidal neurons in the full NGVU model.
    (a) Example voltage traces for 5 randomly chosen active and 5 randomly chosen inactive neurons at $\text{I}_\text{stim}=0.95$\,nA. Shaded time windows are expanded in (c).
    (b) Final state of pyramidal neurons for the simulation shown in (a), active neurons shown in red, silent  neurons are shown in blue.
    (c) Average excitatory (PYR$\rightarrow$PYR) and inhibitory (PV$\rightarrow$PYR) currents in the pyramidal neurons that will remain active (red) or become inactive (blue) for the time windows shaded in (a). 
    The first, shorter, window corresponds to stimulation onset, the second, longer, window demonstrates the critical period just before exit of a fraction of pyramids in DB, it coincides with a transient activation of interneurons, visible from a brief activation of inhibitory synaptic currents. 
    Note the changes in scale to show the currents.
    (d) Separation of active and inactive neurons by the early integral inhibitory and excitatory currents, taken in the first time window shown in (c). Red markers --- active neurons, blue markers --- inactive, dashed line: separation by logistic regression. 
    }
    \label{fig:coupled-1}
\end{figure}

The potassium-uncoupled model displays activity-segregated states for a relatively narrow range of extracellular potassium concentration. It is thus interesting if coupling spiking activity to potassium accumulation in the NGVU model will tend to linger in this fractional activation state.
To test this hypothesis, we return to the full model described by equations (\ref{eq:dV1}--\ref{eq:NA}),
% -- AB я смогу проставить ссылки только после того, как будут все уравнения!
which account for feedback loops linked to activity-dependent dynamics of extracellular potassium, GABA, and Glu, as well as Glu-dependent modulation of noradrenaline level. 
% AB: по-хорошему, все-таки нужен рисунок с растром активности и соответстующими переменными (калий, глутамат, гамк)
In the full model, stimulation of PYR neurons with depolarizing current also leads to general silencing of PV neurons and the dissociation of active and quiescent states in PYR neurons~(Fig.\,\ref{fig:coupled-1}a--b). 
Indeed, in this case stimulation causes rapid firing of PYR neurons and activity-associated accumulation of potassium, providing a positive feedback loop between firing of PYR neurons and potassium-dependent depolarization. This self-drives the system on the verge of DB bifurcation, but the network stabilizes itself due to competition for DB evasion in PYR neurons: the neurons that receive more excitatory synaptic input drop out to DB earlier than others, which reduces the net potassium accumulation rate and the net excitation drive to the rest of the PYR neurons which remain in the rapid firing state just before DB bifurcation. 
This separation is visible both during the first 100\,ms of stimulation and during the critical phase when the dropout occurs~(Fig.\,\ref{fig:coupled-1}c--d). 
There is also moderate prevalence of synaptic inhibition in the surviving neurons in the early stage of the critical period~(Fig.\,\ref{fig:coupled-1}c), but the difference in excitatory drive is the most decisive. 
The separation of neurons according to the integral excitatory and inhibitory currents during the first 100\,ms of stimulation shown in Fig.\,\ref{fig:coupled-1}d resembles the low stimulation case in the uncoupled network shown in Fig.\,\ref{fig:uncoupled}e despite the higher stimulation current in the full model (0.95\,nA in the full model vs 0.3\,nA in the uncoupled model), which in part is due to counteractive effect of the tonic inhibition by extracellular GABA, released by astrocytes due to potassium-dependent depolarization.  

\begin{figure}%[htbp]
  \begin{center}
\resizebox{1\columnwidth}{!}{
\includegraphics{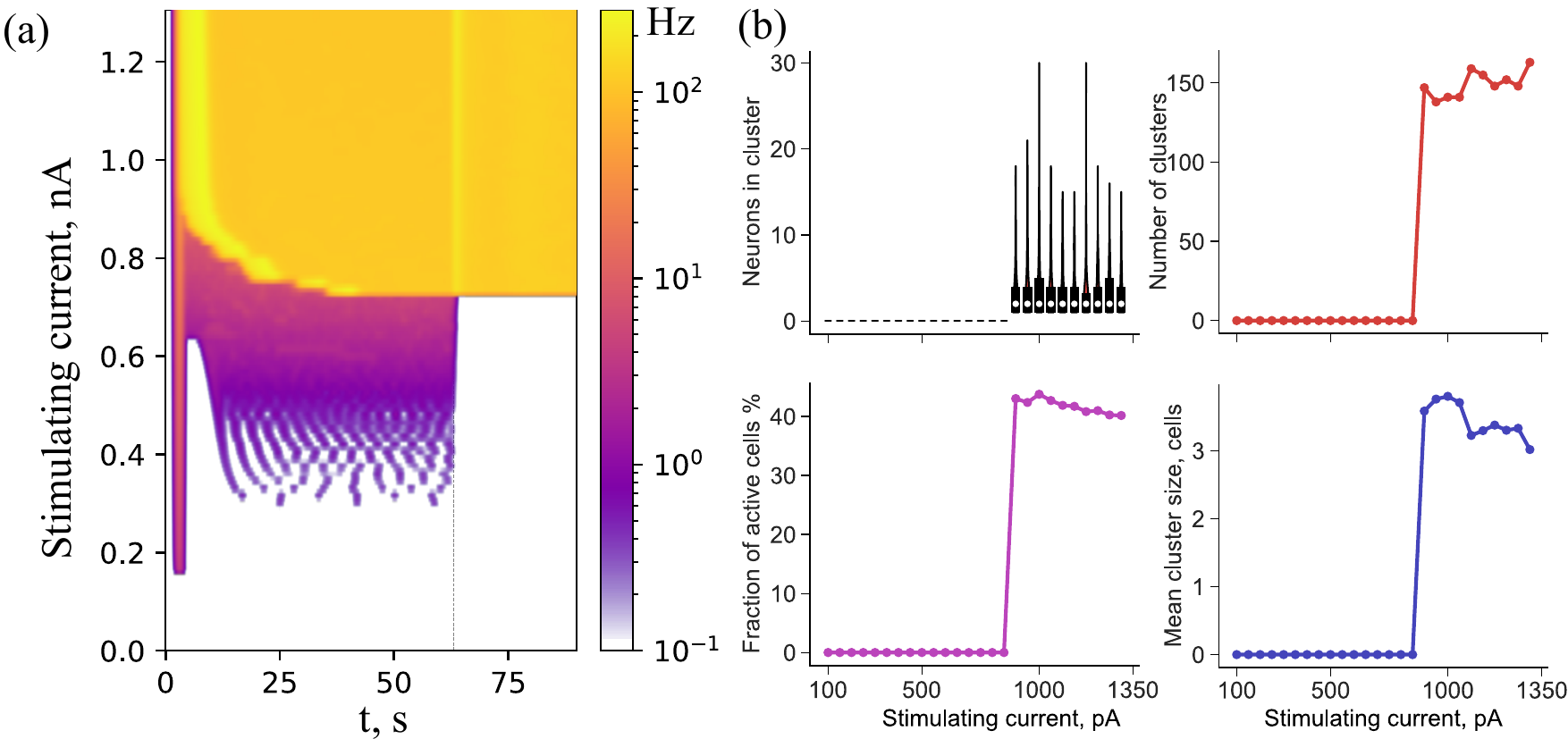}}
 \end{center}
\caption{Influence of stimulation intensity on the spatiotemporal dynamics of network activity
%Влияние интенсивности стимуляции на пространственно-временную динамику активности сети. 
(a) Evolution of network dynamics under varying stimulating current, visualized as average firing rate of pyramidal neurons in each time point. Stimulating current starts at 3\,s and stops at 63\,s, the end of stimulation is marked with a dashed gray line.
%Динамические режимы нейронной сети при различных значениях стимулирующего тока. 
% Хорошо бы и калий также показать как тепловую карту здесь.
(b) Properties of active neural clusters as a function of stimulating current. Panels from left to right: cluster size distribution, total number of active clusters, overall fraction of active cells, and mean cluster size.
}
\label{fig:coupled-Istim}
\end{figure}

We next looked at how the activity of the neuronal network in the NGVU complex depends on depolarization level by systematically varying $I_\text{stim}$~(Fig.\,\ref{fig:coupled-Istim}). 
At low depolarizing current ($I_{\mathrm{stim}} < 300\text{ pA}$), the stimulation is insufficient to initiate sustained excitation. The network response is either entirely absent or limited to a transient response at the onset of stimulation, after which the system returns to the quiescent state.
Gradually raising the current leads to emergence of sustained quasi-periodic collective bursting activity. 
%reflecting bursting of individual pyramidal neurons, enabled by inactivation of the calcium current (Fig.\,\ref{fig:coupled-Istim}a). 
%
Further depolarization leads first to narrowing the inter-burst gaps and then to merging of bursts into homogeneous high-frequency activity. In the distributed system, this regime is characterized by traveling self-sustained waves of higher firing rates, which cross the whole network. 
The network-level bursting regime evolves due to a noradrenaline-dependent negative feedback on PYR neurons: astrocyte depolarization leads to glutamate release~(\ref{eq:glu}), which in turn leads to modulation of noradrenaline (NA) release from \textit{locus coeruleus} axons~(\ref{eq:NA}) and activation of $\alpha_2$ noradrenoreceptors, positively modulating GIRK potassium channels in PYR neurons~(\ref{eq:gna}). In fact, this is the negative feedback on PYR activity that does not depend on inhibitory neurons. Removing this mechanism from the model obliterates the global bursting regime and lowers the threshold for transition to runaway depolarization and dissociated activity. 
Finally, reaching a critical depolarization level of around 750\,pA leads to a sharp, avalanche-like increase in the AP firing rate following a stimulus-dependent delay after the onset of stimulation, driven by rapid accumulation of extracellular potassium ($[K^+]_{e}$) and collapse of interneurons to DB, effectively extinguishing any inhibition in the network. This is followed by a decrease in the mean firing rate due to a fraction of pyramidal neurons dropping out via DB as well. However, a specific heterogeneous fraction of the pyramids forms spatial clusters with pathologically sustained high-frequency activity. Interestingly, this pathological activity remains after the end of external stimulation: removal of stimulating current is associated with a short rise of average firing rate due to more neurons emerging from DB, which, however, are then pushed to DB again due to synaptic excitation from the rest of the network. Under these conditions, the accumulated extracellular potassium level exceeds the astrocytic homeostatic capacity and it sustains high level of excitation in the network, enough to keep a fraction of PYR neurons on the brim of DB.
%This DB is likely due to self-organizing of these neurons into the narrow band of $I_{\text{Ca}^{2+}}$-dependent recovery, demonstrated in Fig.\,\ref{fig2}.

The formation of the cluster structure resembles a threshold phase transition governed by the tonic stimulus intensity (Fig.\,\ref{fig:coupled-Istim}b). Below the critical threshold current ($\sim 750$\,pA), the network state remains spatially homogeneous. Exceeding the threshold value leads to an abrupt dissociation of the neuronal population into an ensemble of approximately 150 stable microclusters, demonstrating a strong non-linearity in the interaction between membrane currents and the dynamics of extracellular ion homeostasis.

Thus, our simulations at different depolarization levels predict that the increase in the metabolic and electrical load on the tissue pushes the system from a regime of regular responses toward complex self-organization and functional activity suppression via mechanisms involving the microenvironment of neurons.

\begin{figure}%[htbp]
  \begin{center}
\resizebox{1\columnwidth}{!}{
\includegraphics{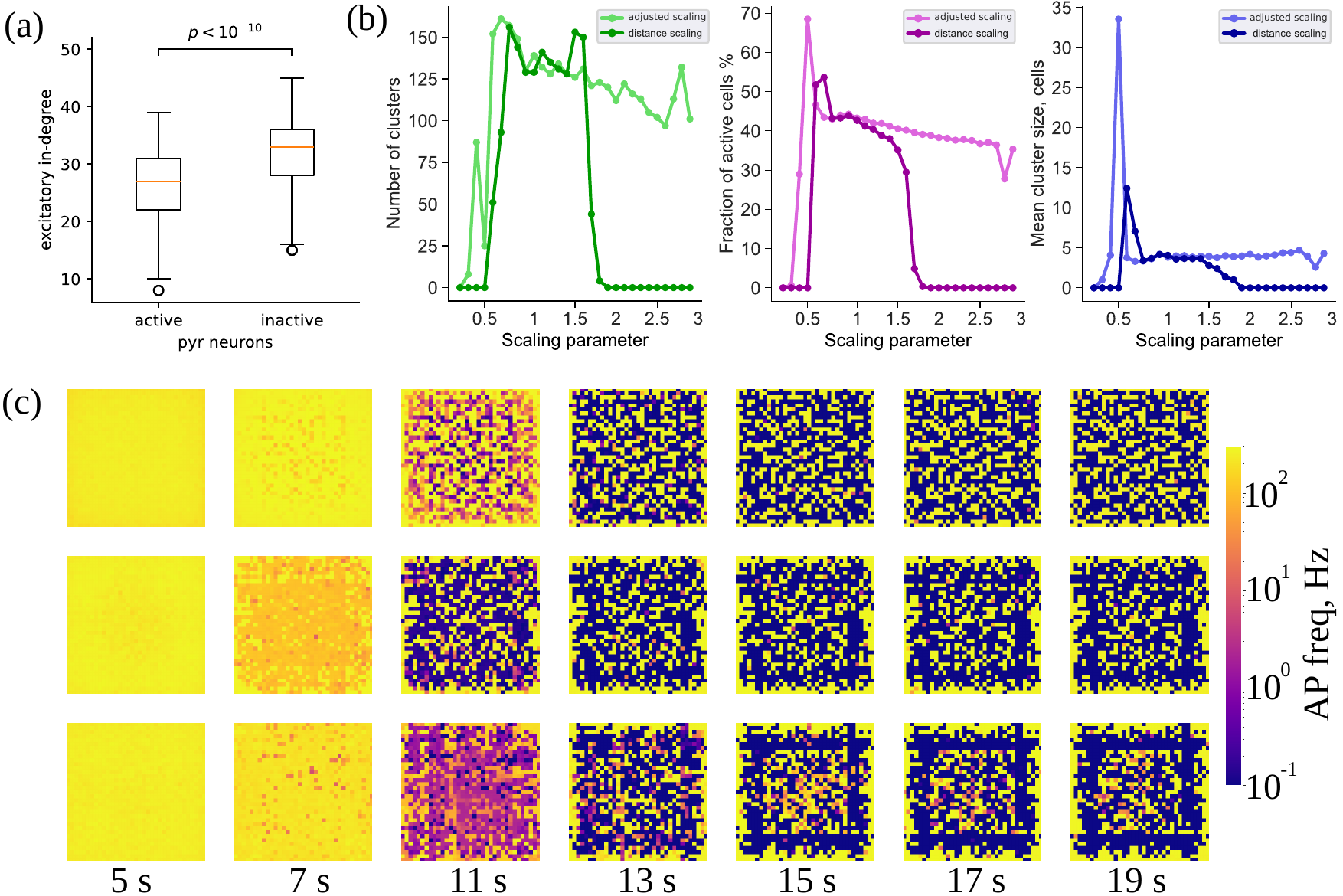}}
 \end{center}
\caption{Spatiotemporal activity patterns, network fragmentation, and the role of synaptic connectivity. 
(a) Comparison of excitatory in-degree distributions between active and inactive pyramidal neuron populations.
(b) Effect of connectivity parameters on cluster characteristics as a function of the scaling factor: distance scaling alters only the connectivity radius. Panels from left to right: total number of active clusters, overall fraction of active cells, and mean cluster size. 
% ** AB: легенды очень мелкие. если есть возможность, стоит увеличить шрифт в 1.5 раза
% AB: желательно, если есть возможность, поменять цвета местами: полный scaling сделать темным (и вторым в легенде), а только расстояния -- светлым (и первым в легенде)
(c) Evolution of spatial patterns of pyramidal neuron firing rates across different simulations with varying connectivity scales at different time points;
the connectivity radius increases (scales $\kappa=1$, $\kappa=2$, and $\kappa=3$, top-to-bottom).
% *** AB: нужно на рисунке указать для каждой строки в (c) какие параметры связности использовались; указал, но стоит перепроверить.
% *** AB: это случай полного масштабирования же?
}
\label{fig:coupled-connectivity}
\end{figure}

%Based on the biophysical characteristics of isolated cells, we investigated the spatiotemporal dynamics of a neuronal population under prolonged stimulation. 
%What determines whether activity of a given pyramidal neuron will be quenched or maintained? 
%The escape from DB depends on the depolarization level, which in the model has two origins: the applied stimulating current and the excitatory postsynaptic currents. Because the stimulating current is spatially homogeneous, we hypothesized that the firing heterogeneity should result from the synaptic connectivity pattern for each neuron.

Above we show that the final activity outcome of PYR neurons depends on the relative accumulated inhibitory and excitatory synaptic currents in the first 100\,ms after stimulation onset and later during the critical stage of transition to the pathological activity~(Fig. \ref{fig:coupled-1}).
Indeed, the cells that remain active have, on average, fewer excitatory inputs than the neurons that are silenced (Fig.\,\ref{fig:coupled-connectivity}a). This however only partly explains the dissociation between the active and inactive neurons, and the remaining part is likely to emerge from race conditions of which neuron first drives its neighbors to DB during the phase of initial rapid extracellular $\mathrm{K}^+$ accumulation. Still, neurons with a lower number of incoming excitatory connections receive a smaller total influx of recurrent excitation, which helps them to avoid entering the terminal DB and remain within the range of stable firing or secondary recovery. Continuous AP generation is accompanied by the local accumulation of extracellular $\mathrm{K}^+$ ions and glutamate, which selectively drives surrounding neurons with higher connectivity into a state of deep DB, thereby forming the stable spatially fragmented activity pattern.

%The ability of some pyramidal neurons to maintain spiking activity under pathological hyperexcitation is determined by its position within the network's synaptic topology, specifically by its in-degree of incoming excitatory connections (see $k_{\text{in}}$ in-degree in Fig.\,\ref{fig:coupled-connectivity}a). Distribution analysis revealed that cells maintaining regular firing throughout the stimulation period are characterized by a lower median $k_{\text{in}}$ value compared to neurons transitioning into an inactive state. 

%Topological heterogeneity underlies the mechanism of local activity suppression mediated by the extracellular environment. Neurons with a low number of incoming connections receive a smaller total influx of recurrent excitation, which prevents them from entering the terminal DB. Maintaining their membrane AP within the range of stable generation or secondary reactivation, these cells function as pacemakers. Continuous AP generation is accompanied by the local accumulation of extracellular $\mathrm{K}^+$ ions and glutamate, which selectively drives surrounding neurons with higher connectivity into a state of deep DB, thereby forming a stable spatially fragmented activity pattern.

If network topology defines the heterogeneity of maintained activity, changing connectivity rules should transform the emerging spatial patterns. In our model synapses are formed stochastically with a probability which decays exponentially with the squared distance from the presynaptic neuron. Increasing just the distance constant $\sigma = \kappa\sigma_1$ of this decay, where $\sigma_1=7$\,$\mu$m and $\kappa\in[0.2,\dots 3,]$ would lead in the case $\kappa > 1$ to a more densely connected network with a larger amount of excitatory and inhibitory synapses. When $\kappa < 1$, the network becomes more sparsely connected. We refer to this connectivity variation as ``distance scaling'' below. To adjust for the increased or diminished connectivity, one can reduce or increase the maximal conductance of all synapses to keep the amount of excitation constant, which we below refer to as ``adjusted scaling''.  

In the case of unadjusted distance scaling the clusters appear at $\kappa=0.5$, probably because below this value the network is too sparse to support any kind of spatial self-organization, whereas doubling the length constant of connectivity ($\kappa \gtrsim 2$) leads to a sharp decline in the number of clusters, fraction of active cells, and mean cluster size~(Fig.\,\ref{fig:coupled-connectivity}b). The loss of the ability to form active clusters at unadjusted high connectivity radius reflects the increased amount of excitation in the network, pushing pyramidal neurons to DB beyond the possibility for recovery. 
%Under isometric scaling of the connection radius alone (distance scaling), which dilutes local synaptic density, the number of active clusters and the fraction of firing cells monotonically decrease to zero (Fig.\,\ref{fig:coupled-connectivity}b).  
Under adjusted scaling, the clusters start forming at smaller $\kappa$ values and the clustered state remains parametrically stable at large $\kappa$ because the enlarged connection radius is balanced by synaptic weight renormalization to preserve overall excitation.
Over a wide scaling range, the fraction of active neurons stabilizes at $\sim 40\%$, with the system self-segregating into many small clusters (an average size of $3\text{--}4$ cells).

%The pattern of spatiotemporal evolution is determined by the spatial radius of synaptic connectivity (Fig.\,\ref{fig:coupled-connectivity}c). As the connectivity radius increases, network clustering is observed: stable foci of highly active neurons emerge, surrounded by inactive elements. Higher connection density leads to a more homogeneous solution due to the diffuse accumulation of extracellular potassium across a major part of the neuronal population.
The forming spatial pattern of active and inactive neurons is defined by the length constant of synaptic connectivity in the network (Fig.\,\ref{fig:coupled-connectivity}c). As the length constant increases from 7\,$\mu$m in the top row to 21\,$\mu$m in the bottom row, there appears more spatial order, mostly due to boundary effects. The neurons at the edge of the modeled NGVU grid will receive significantly less excitatory input at long-distance connectivity because they lack the neighbors outside the grid. Consequently, these neurons are more likely to win the competition for DB avoidance and they successively suppress their immediate neighbors inside the grid. Suppression of activity in the interior of the grid allows for the pattern formation in the center of the modeled space. 
%
%-- AB: Станислав не согласен с формулировкой ниже, утверждает, что там не промежуточные
%       частоты активности, а переключения между максимальной частотой и молчанием.
%       стоит перепроверить.
Interestingly, the pattern at short-distance connectivity appears almost binary: some neurons fire at maximal rate, while others are completely silent. On the other hand, the increase in the connectivity distance leads to more intermediate average firing rate states, as seen from comparison of the first and bottom rows of Fig.\,\ref{fig:coupled-connectivity}c. 
% -- Добавил это предложение.
This is a result of a fraction of neurons fluctuating between brief periods of high-frequency spiking and quiescence, which after temporal smoothing appears as intermediate frequency firing.

\subsection{Spatiotemporal self-organization of activity under local system perturbations}
%{Пространственно-временная самоорганизация активности при локальных возмущениях системы}

\begin{figure}%[htbp]
  \begin{center}
\resizebox{1\columnwidth}{!}{
\includegraphics{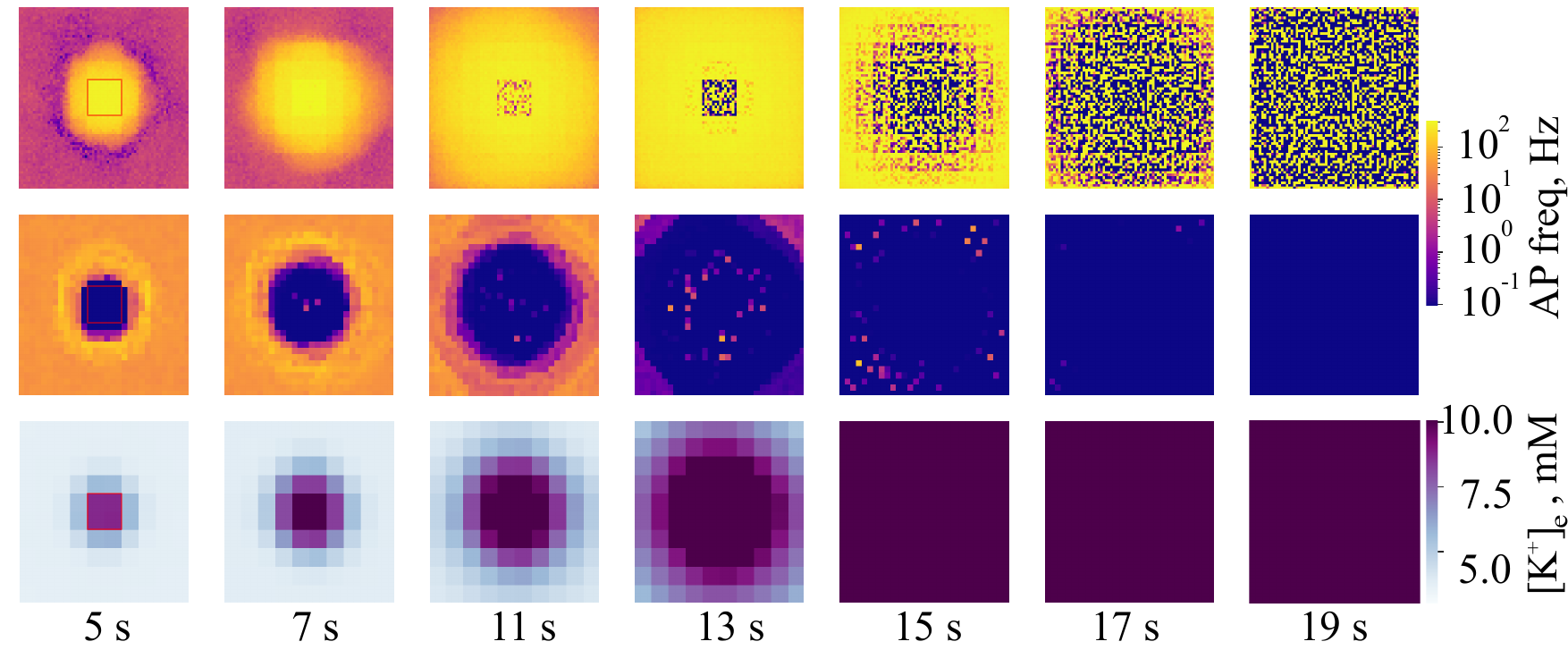}}
 \end{center}
\caption{Spatiotemporal patterns of neuronal firing and extracellular potassium concentration under localized NGVU parameter modifications: an increase in stimulating current in the central $2 \times 2$ area to 1600 pA (800 pA in the rest of the network). The top, middle, and bottom panels show the firing rates of pyramidal neurons and interneurons, and the extracellular $\mathrm{K}^+$ concentration $[\mathrm{K}^+]_{\mathrm{ext}}$, respectively.
%Верхние графики соответствуют частоте firing rate пирамидальных нейронов, средние --- интернейронов, нижние --- концентрации $[\mathrm{K}^+]_{\mathrm{ext}}$
}
\label{fig5}
\end{figure}

% -- AB: remember the FBI :) 
%This section investigates the stability of the developed NGVU model against local pathological perturbations. 
The simulations above were run in spatially homogeneous settings, affecting the whole NGVU grid in the same way. It is also interesting to understand how local pathological perturbations in a single NGVU would affect behavior of the whole ensemble.
To this end, through a series of simulations, we examine regimes characterized by focal hyperexcitability or extracellular potassium clearance failure (Figs.~\ref{fig5}\text{--}\ref{fig7}). Three qualitative scenarios are analyzed, tracking the dynamics of the firing rate of pyramidal neurons (PYR) and PV interneurons, and extracellular potassium concentration $[\mathrm{K}^+]_{\mathrm{ext}}$ for each case.

%В данном разделе проведено исследование устойчивости разработанной модели NGVU к локальным патологическим воздействиям. На основе серии численных экспериментов исследованы режимы с повышенной возбудимостью и с локальной недостаточностью  гомеостаза внеклеточного калия (see Fig.\,\ref{fig5}---\ref{fig7}). Рассмотрены три качественных сценария, для каждого из которых фиксировалась динамика частоты импульсации пирамидальных нейронов (PYR), PV-интернейронов и концентрации внеклеточного калия $[\mathrm{K}^+]_{\mathrm{ext}}$.

To simulate a local increase in stimulation intensity (Fig.~\ref{fig5}), the stimulation current within a central patch of the size $2 \times 2$ NGVUs was doubled to $1600\text{ pA}$ compared to $800\text{ pA}$ in the rest of the network.
Within the first seconds the hyperexcited focus is formed, characterized by the high spiking rate of the pyramidal neurons, quenched activity of PV neurons and local accumulation of extracellular potassium $[\mathrm{K}^+]_{\mathrm{e}}$. The zone of suppressed interneuron activity continues to grow following the expansion of the area with elevated $[\mathrm{K}^+]_{\mathrm{e}}$.  However, by the $11^{\text{th}}$ second, a fraction of pyramidal neurons in the overstimulated focus start entering the DB state, which is seen as a spatially heterogeneous drop in firing rate of the pyramidal neurons. Gradually, this sequence of events, that is, a rapid increase in firing rate of the pyramidal neurons followed by segregation into silent and active neurons engulfs the whole system, by which time all interneurons are silent and $[\mathrm{K}^+]_{\mathrm{e}}$ is pathologically elevated in each NGVU. In this scenario, a focus with abnormal excitatory input can trigger the conversion of the neighboring NGVUs into pathological activity state via horizontal excitatory synaptic connections, accumulation of potassium, and suppressing the inhibition in the network due to higher susceptibility of interneurons to entering DB.

\begin{figure}%[htbp]
  \begin{center}
\resizebox{1\columnwidth}{!}{
\includegraphics{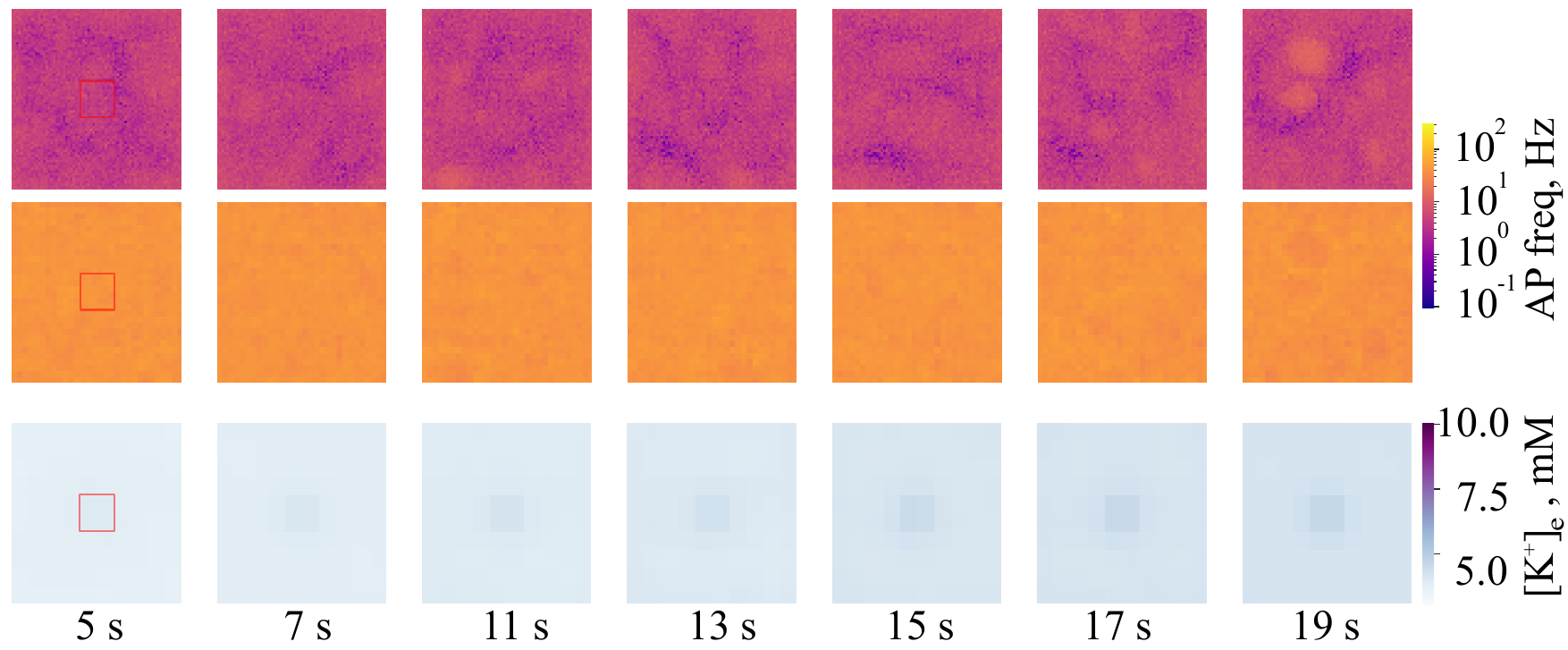}}
 \end{center}
\caption{Spatiotemporal patterns of neuronal firing and extracellular potassium concentration under localized NGVU parameter modifications: $10\times$ reduction of potassium clearance rate in the central $2 \times 2$ area. The top, middle, and bottom panels show the firing rates of pyramidal neurons and interneurons, and the extracellular $\mathrm{K}^+$ concentration $[\mathrm{K}^+]_{\mathrm{e}}$, respectively.
}
%Верхние графики соответствуют частоте firing rate пирамидальных нейронов, средние --- интернейронов, нижние --- концентрации $[\mathrm{K}^+]_{\mathrm{ext}}
\label{fig6}
\end{figure}

To analyze the effect of impaired local potassium clearance, we compromised  potassium clearance capacity by increasing the corresponding equilibration time constant, $\tau_K$, tenfold in the units within central $2 \times 2$ (Fig.\,\ref{fig6}) and $6 \times 6$  (Fig.\,\ref{fig7}) regions of the modeling space.
% *** AB: уравнение для K+ нужно вставить в описание модели, чтобы там был параметр $\tau_K$, на который я здесь ссылаюсь. И нужно добавить label для уравнения, чтобы можно было здесь вставить ссылку на него. 
%
%potassium clearance rate tenfold in the unites within central $2 \times 2$ (Fig.\,\ref{fig6}) and $6 \times 6$  (Fig.\,\ref{fig7}) domains with astrocyte-mediated extracellular potassium clearance reduced to near-zero values. 
%
In the case of the small affected domain ($2 \times 2$), the system is capable to compensate for the local impairment of the potassium clearance by diffusive exchange with the neighboring units  
%demonstrates an adaptive capability 
(Fig.\,\ref{fig6}). In this case, extracellular potassium (bottom row) is siphoned outside the pathological focus thanks to exchange with the adjacent functional astrocytes and efficient uptake by them, while the neuronal firing rate remains spatially uniform. 

% TODO: try 4x4 pathological zone
Expanding the impaired zone to $6 \times 6$ units (Fig.\,\ref{fig7}) leads to the breakdown of the compensatory mechanism: the center of the domain exhibits the progressive accumulation of $[\mathrm{K}^+]_{\mathrm{e}}$ exceeding the physiological threshold of $10\text{ mM}$. Within this region 
%both pyramidal neurons and  %% AB: ну нет же!
interneurons switch into the DB regime, allowing runaway excitation and formation of the stable pathological structure, featuring elevated high-potassium and segregation of pyramidal neurons into rapidly spiking and silent. Comparison of the two cases suggests an existence of a critical spatial threshold for the metabolic defect, below which the homeostatic reserves of the NGVU are able to fully compensate for the local insufficiency.

Thus, the results presented in Figs.~\ref{fig5}--\ref{fig7} demonstrate the critical role of astrocytic network spatial connectivity and potassium diffusion mechanisms. On the one hand, the system maintains stability against local perturbations, which are successfully counteracted within a specific spatial scale. On the other hand, exceeding this threshold scale or locally increasing the stimulation intensity leads to pathological activity patterns due to competitive escape from DB. In this case, the system transitions into regimes that can be linked to neuropathological states such as spreading depolarization or epileptiform activity.

%Таким образом, результаты, представленные на Fig.\,\ref{fig5}--\ref{fig7}, демонстрируют ключевую роль пространственной связности астроцитарной сети и механизмов диффузии калия. С одной стороны, система демонстрирует устойчивость к локальным возмущениям, которые успешно нивелируются в пределах определенного пространственного масштаба. С другой стороны, превышение этого порогового масштаба или локальное увеличение интенсивности стимула приводит к потере возбудимости ткани посредством ДБ. В последнем случае система переходит в режимы, характерные для таких патологических состояний, как распространяющаяся деполяризация (spreading depolarization) или эпилептиформная активность.

\begin{figure}%[htbp]
  \begin{center}
\resizebox{1\columnwidth}{!}{
\includegraphics{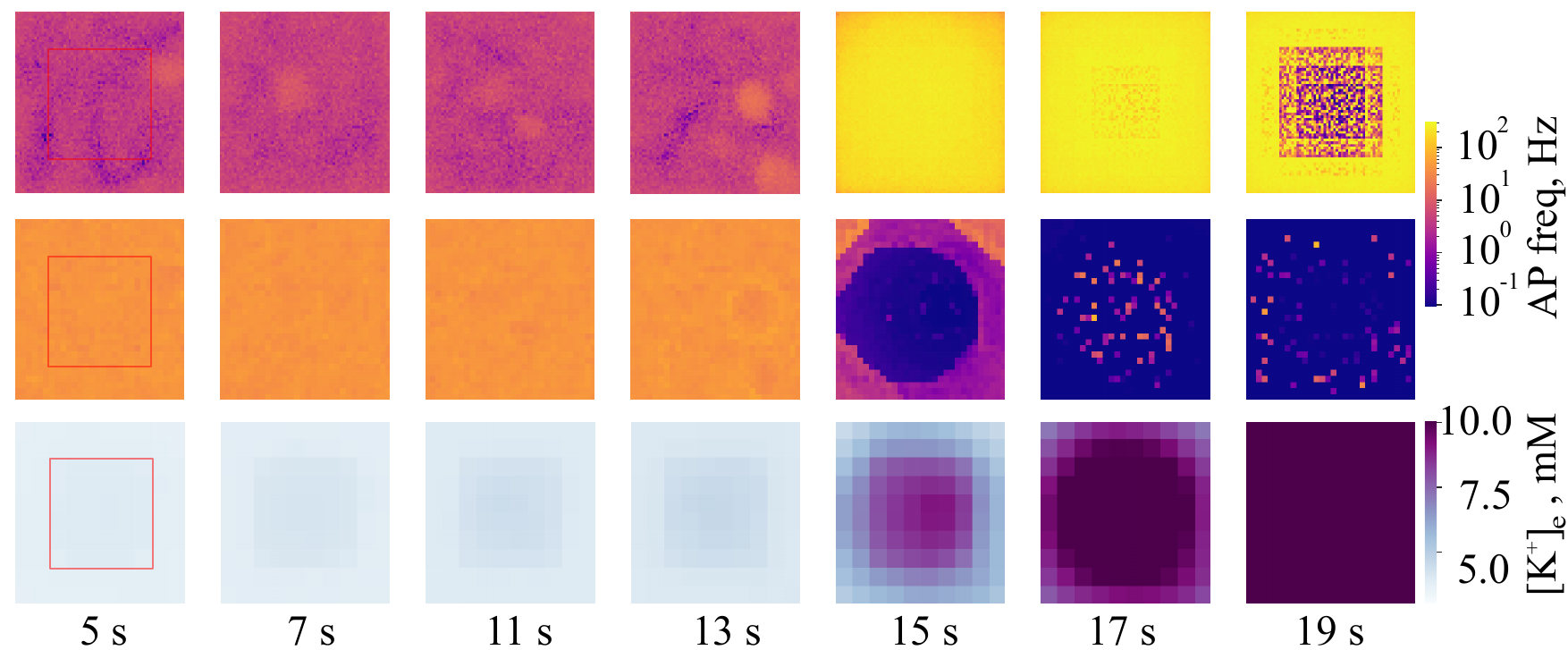}}
 \end{center}
\caption{Spatiotemporal patterns of neuronal firing and extracellular potassium concentration under localized NGVU parameter modifications: (c) reduction of potassium clearance rate to near-zero values in the $6 \times 6$ area.
The top, middle, and bottom panels show the firing rates of pyramidal neurons and interneurons, and the extracellular $\mathrm{K}^+$ concentration $[\mathrm{K}^+]_{\mathrm{ext}}$, respectively.}
%Верхние графики соответсвуют частоте firing rate пирамидальных нейронов, средние --- интернейронов, нижние --- концентрации $[\mathrm{K}^+]_{\mathrm{ext}}$}
\label{fig7}
\end{figure}

%%%%%%%%%%%%%%%%%%%%%%%%%%%%%%%%%%%%%%%%%%%%%%%%%%%%%%%%%%%%%%%%%%%%%%%%%%%%%%%%

\section{Discussion and Conclusions}
%План
\begin{comment}
- [X] Succeptibility  of PV vs PYR
- [X] Synaptically defined dissociation of activity levels in uncoupled network
- [X] Self-organization in coupled network
- [ ] Back to the effect of synaptic connectivity
- [ ] Effects of spatial heterogeneity

\end{comment}

We present a biophysically detailed spatially extended NGVU model as a tool for the analysis of complex spatiotemporal dynamics in neural tissue. The model incorporates electrical activity of excitatory and inhibitory synaptically connected neurons, astrocyte-mediated extracellular ion exchange processes, and extrasynaptic neurotransmitter uptake and spillover. Here we use the suggested model to explore pathological excitability regimes linked to activity-dependent accumulation of extracellular potassium.

We start with observing that Hodgkin-Huxley models of fast-spiking interneurons have lower thresholds to enter the state of depolarization block by either applying depolarizing current or increasing extracellular potassium concentration [K$^+$]$_\text{e}$ than models of excitatory pyramidal neurons. This is supported by experimental evidence in \textit{in vitro} models of epilepsy where pyramidal neurons show only brief DB episodes and maintain high-frequency spiking throughout the ictal discharge, whereas interneurons remain locked in prolonged DB state~\cite{ziburkus2006interneuron}.

%One of the key observations of the study is the identified difference in DB thresholds between neuron types arising from their distinct ion channel compositions and current-voltage characteristics. Inhibitory interneurons transition to DB at significantly lower values of the stimulating current and extracellular potassium concentration $[\mathrm{K}^+]_{\mathrm{ext}}$ compared to pyramidal neurons. 
Thus, early inactivation of inhibitory units leads to a local shift in the excitation-inhibition ($E/I$) balance toward activation, aligning with recent experimental and computational evidence on the modulatory role of GABAergic populations in pathological network dynamics~\cite{baspinar2025gabaergic}. This mechanism can act as a primary trigger of network hyperexcitability, accompanying propagation of pathological regimes in neural tissue. 
%%  -- obsolete: In accordance with the model architecture, unlike interneurons, pyramidal neurons feature voltage-gated calcium currents ($I_{\mathrm{Ca}}$), which ensure their ability to recover spike activity under sustained depolarization. Due to the slow inactivation of $I_{\mathrm{Ca}}$, the membrane potential decreases during strong depolarization, resulting in the reactivation of sodium channels. This allows the cell to maintain AP generation even under conditions of high extracellular potassium.

%The intrinsic $I_{Ca}$-mediated DB escape mechanisms identified in this modeling study are fully consistent with experimental recordings of in vitro seizure-like events, where pyramidal neurons show only brief DB episodes and maintain high-frequency spiking throughout the ictal discharge, whereas interneurons remain locked in prolonged DB~\cite{ziburkus2006interneuron}. In contrast, interneurons lacking slow recuperative currents exit DB exclusively in a passive manner during late stages of the discharge, once astrocytic clearance and diffusion reduce $[K^+]_{e}$ and membrane depolarization below the bifurcation threshold.

A potassium-uncoupled synaptically connected network with $[\mathrm{K}^+]_{\mathrm{e}}$ as a parameter displays dissociation of actively firing and DB-silenced pyramidal neurons for a range of I$_\text{stim}$ and [K$^+$]$_\text{e}$ combinations instead of global silencing of activity.  This happens because of heterogeneity of excitatory and inhibitory synaptic inputs during the stimulus onset, not all pyramidal neurons cross the DB bifurcation boundary initially, but increase the probability of pushing their postsynaptic targets to DB by effectively increasing their combined depolarization current. Coupling of neuronal spiking activity to [K$^+$]$_\text{e}$ accumulation and astrocyte-dependent clearance in the NGVU model shows that stimulation of pyramidal neurons by depolarizing current can lead to self-organization of the network to this fractionally active state. A positive feedback loop between potassium accumulation and increased excitability first pushes interneurons to DB allowing for runaway excitation which brings pyramidal neurons on the edge of DB, where a new quasi-steady state  forms: progressive dropout of pyramids to DB gradually attenuates potassium accumulation, balancing it with astrocyte-dependent clearance and reducing the depolarization pressure for the neurons that are still active due to termination of excitatory input to them. In the suggested model we also account for tonic inhibition via GABA released from astrocytes due to potassium-driven reversal of GAT transporter, which has a stabilizing effect on the network by raising the I$_\text{stim}$ threshold to enter the runaway excitation state. This stabilizing effect is combined with a different negative feedback described in the model: here, glutamate spillover due to synaptic activity positively modulates release of noradrenaline from \textit{locus coeruleus} axons~\cite{Paukert:2014:Norepinephrine} which activates GIRK potassium current via $\alpha_2$ adrenoreceptors. We should note that the effect of noradrenaline neuromodulation on pyramidal neurons can be more complex, involving $\alpha_1$ and $\beta_{1,2}$ adrenoreceptors, coupled to production of IP$_3$ or adenylate cyclase with distinct downstream mechanisms, correspondingly, both positively and negatively modulating neuronal excitability. Here we only add a, possibly, oversimplified view on this by only incorporating the negative, $\alpha_2$-based feedback loop and leave a more detailed exploration of noradrenaline effects on the slip to pathological excitation regimes to future studies. 
%- GABA, NA effects

%Neuronal transmembrane currents and extracellular potassium dynamics form a feedback loop, with ionic accumulation directly governing the transition into DB. 
%Depending on network topology, local interactions scale into distinct macroscopic regimes. 
%At short synaptic connectivity radii, the network exhibits clustered activity, 
%whereas high connection density combined with potassium diffusion can trigger global DB, which is characteristic of spreading depolarization states. 
% -- AB: последнего не увидел в результатах!

The separation of pyramidal neurons to active and silenced is defined by their synaptic connectivity properties: on the one hand, active neurons have a lower number of incoming excitatory synapses and on the other, the segregation depends on the net inhibitory input in the short temporal window when the interneurons are still active. 
By varying the characteristic length of synaptic connectivity we observed how local interactions scale into distinct macroscopic regimes. 
Making the network more globally connected by increasing the connectivity radius leads to appearance of more spatial order in the clusters of active neurons, in part due to edge effects; additionally, at longer synaptic connections we observe transient switching between actively firing and silent states in a fraction of neurons.

Within the locally clustered regime, the system exhibits local competition between neurons for maintaining the active state. Neurons with a low in-degree demonstrate higher resistance to DB due to a reduced recurrent excitatory load from the surrounding tissue. 
%Transitioning into persistent isolated active cells within a silenced network background, such cells act as autonomous pacemakers. 
% -- AB:  это не пейсмейкеры...
These active neurons continue to exert the sustained depolarizing pressure on their neighbors continuously generating APs and releasing $\mathrm{K}^+$ ions and adding to glutamate spillover. Conversely, the already silenced surrounding cells stop providing excitatory input to the active neurons, thus preventing them from entering DB. This mutual recurrent feedback loop stabilizes the active core and drives the formation of robust, spatially fragmented network activity patterns.

The considered effects highlight critical mechanisms governing the transition into DB and recovery from it within spatially distributed neurogliovascular networks. Specifically, the analysis of local metabolic disturbances reveals a high buffering capacity of the tissue: small local potassium clearance defects are successfully compensated by diffusion and spatial buffering from neighboring healthy NGVUs, whereas exceeding the critical spatial scale of the damaged region leads to a breakdown of ionic homeostasis, macroscopic network fragmentation, and the formation of persistent pathological structures.

The pathological regimes observed in the studied system can become self-sustaining in the absence of a continuous external driver. Even after the external depolarizing force is stopped, the system can remain in the state of spatially segregated activity. This predicts that seizure-like activity and other hyperexcitability pathologies associated with activity-dependent potassium build-up, for example in peri-ischemic penumbra, may not lead to full depolarization block, but to self-sustained high-frequency firing of a fraction of neurons, prolonging elevated metabolic demand on the tissue and potentially aggravating the pathological defects. 
%or the local ionic clearance defect is corrected, the system relaxes to the normal physiological state. This emphasizes the fundamental role of metabolic support systems in stabilizing network activity and preventing hyperexcitability.

In conclusion, the developed two-dimensional NGVU model confirms the hypothesis that local anomalies in neuronal activity can be sustained through a complex interplay between intrinsic neuronal excitability, astrocytic ion homeostasis, and the topological structure of the network.
Future development of the model can be aimed at incorporating explicit vascular dynamics, where vasoactive metabolites released in response to pyramidal neuron activity and astrocytic $Ca^{2+}$ signaling modulate vascular tone and local blood perfusion, as well as a more detailed description of noradrenergic neuromodulation within the framework of full neurogliovascular coupling.

\label{app1}

\section{Acknowledgments}
%The study was conducted under the state assignment of Lomonosov Moscow State University.
This work was supported by RSF grant 23-74-00006.%? (to SVL, NAB and ARB)

\bibliographystyle{elsarticle-num} 



%\begin{thebibliography}{00}

%% For numbered reference style

%\bibitem{lamport94}
%  \textit{\LaTeX: a document preparation system},
%  Addison Wesley, Massachusetts,

%\end{thebibliography}

\bibliography{astrocytes.bib}{} 
\end{document}